\documentstyle[12pt,aaspp4]{article}

\def\phihat{\hat \varphi}
\def\zhat{\hat z}
\def\rma{{\rm A}}
\def\vphi{v_\varphi}
\def\uphi{u_\varphi}
\def\bphi{B_\varphi}
\def\ro{R_0}
\def\ra{R_{\rm A}}
\def\rl{R_1}
\def\vpone{v_{\rm p,\, 1}}
\def\rd{R_{\rm D}}
\def\rw{R_{\rm W}}
\def\bv{{\bf v}}
\def\bu{{\bf u}}
\def\bB{{\bf B}}
\def\vp{v_{\rm p}}
\def\Bp{B_{\rm p}}
\def\va{v_{\rm A}}
\def\vf{v_{\rm f}}
\def\cs{c_{\rm s}}
\def\vaphi{v_{\rm A, \varphi}}
\def\vap{v_{\rm A, p}}
\def\alf{Alfv\'en}
\def\ma{M_{\rm A}}
\def\ams{M_{\rm A}^{-2}}

\def\mf{M_{\rm F}}
\def\cale{{\cal E}}
\def\kms{{\rm\ km\,s^{-1}}}
\def\cm{{\rm\ cm}}
\def\yr{{\rm\ yr}}
\def\au{{\rm\ AU}}
\def\msun{{\rm M}_\odot}

\lefthead{Ostriker}
\righthead{Cylindrical Disk Winds}

\begin{document}

\title{Self-Similar Magnetocentrifugal Disk Winds with Cylindrical Asymptotics}

\author{Eve C. Ostriker}
\affil{Harvard-Smithsonian Center for Astrophysics, 60 Garden St., Cambridge, MA 02138}
\centerline{and}
\affil{Department of Astronomy \\
University of Maryland, College Park MD 20742; 
ostriker@astro.umd.edu (present address)
}

\bigskip

\centerline{\bf ABSTRACT}

We construct a two-parameter family of models for self-collimated,
magnetized outflows from accretion disks.  As in previous
magnetocentrifugal wind solutions, a flow at zero initial poloidal
speed leaves the surface of a disk in Kepler rotation about a central
star, and is accelerated and redirected toward the pole by rotating,
helical magnetic fields which thread the disk.  At large distances
from the disk, the flow streamlines asymptote to wrap around the
surfaces of nested cylinders, with velocity $\bf v$ and magnetic field
$\bf B$ directed in the axial ($\hat z$) and toroidal ($\hat \varphi$)
directions.  In the asymptotic regime, the velocity secularly decreases
with cylindrical radius $R$ from the inside to the outside of the flow
because successive streamlines originate in the circumstellar disk in
successively shallower portions of the stellar potential.  In
contrast to previous disk wind modeling, we have explicitly
implemented the cylindrical asymptotic boundary condition to examine
the consequences for flow dynamics.  The present solutions are
developed within the context of $r$-self-similar flows, such that $\bf
v$, the density $\rho$, and $\bf B$ scale with spherical radius $r$ as
${\bf v}\propto r^{-1/2}$, $\rho\propto r^{-q}$, and ${\bf B}\propto
r^{-(1+q)/2}$; $q$ must be smaller that unity in order to achieve
cylindrical collimation.  We self-consistently obtain the shapes of
magnetic field lines and the $\theta$-dependence of all flow
quantities.  The solutions are characterized by $q$ together with the
ratios $\ra/\rl$ and $\ro/\rl$, where for a given streamline $\ro$ is
the radius of its footpoint in the disk, $\ra$ is the cylindrical
radius where the flow makes an \alf\ transition, and $\rl$ is its
final asymptotic cylindrical radius.  For given $q$ and $\ro/\rl$,
$\ra/\rl$ must be found as an eigenvalue such that the
\alf\ transition is made smoothly.  In the solutions we have found,
the asymptotic poloidal speed $v_z$ on any streamline is typically
just a few tenths of the Kepler speed $\Omega \ro$ at the
corresponding disk footpoint, while the asymptotic rotation speed
$v_\varphi$ may be a few tenths to several tenths of $\Omega\ro$.  The
asymptotic toroidal \alf\ speed $\vaphi = \bphi/\sqrt{4\pi\rho}$ is,
however, a few times $\Omega\ro$; thus the outflows remain
magnetically dominated, never making a fast-MHD transition.  We
discuss the implications of these models for interpretations of
observed optical jets and molecular outflows from young stellar systems.  
We suggest that the difficulty of achieving strong collimation in vector
velocity simultaneously with a final speed comparable to $\Omega \ro$
argues against isolated jets and in favor of models with broader
winds.


\section{Introduction}

\subsection{Observational Context}

Energetic, collimated jets and outflows are produced by accreting systems in 
a wide range of astronomical environments, from young stars 
forming in cold molecular gas clouds, to white dwarfs, 
neutron stars, and black holes in evolved binaries, to 
supermassive black holes within active galactic nuclei 
(e.g. \cite{liv97}).  Radiation pressure, thermal
pressure, and magnetic stresses may all play a role in accelerating 
and collimating such
flows.  For the case of low-mass pre-main-sequence (PMS) stars, the 
high mass and momentum losses observed (\cite{lad85},\cite{edw93},
\cite{fuk93}, \cite{bac96}) 
have led theorists to believe that magnetic forces are the 
essential ingredient in driving these winds (e.g. \cite{kon93},
\cite{shu96}).  The densities and velocities 
involved place such flows from young stars in the regime of
nonrelativistic magnetohydrodynamics (MHD).  The prevalence of jets
and outflows from pre-main-sequence stars argues that they are an
inevitable byproduct of star formation; outflows may in fact help
determine the eventual mass of the star that forms (\cite{shu87}), and
also have a profound influence on the dynamical evolution of the
parent cloud (\cite{nor80}).  Thus, understanding the physics of
cold, nonrelativistic MHD winds from accreting systems 
is crucial for modeling of
star formation, and also informative for studies of other systems where
magnetic fields are probably important in driving and confining winds
-- but where electromagnetic, thermal, radiative, and relativistic
effects may complicate the dynamics (e.g. \cite{beg84}, \cite{bla90}).

A topic of great theoretical and observational
interest in studies of PMS stars is the dynamical connection between the 
fast, ionized jets seen near the presumed polar axis of the wind,
and the slower and less-collimated molecular outflows that surround them.
Molecular outflows are sometimes ``jetlike'' in appearance
(\cite{bac92}) but often more poorly confined spatially, while 
still having strongly directed bipolar momenta (e.g. \cite{lad96}).
Large outflow masses and red/blue lobe asymmetries argue 
that the bulk of the material in 
molecular outflows is swept up from the surrounding cloud rather than itself
comprising a wind, but there is still no definitive model for the 
process that imparts momentum to the cloud. 
The observed line-of-sight-velocity/plane-of-sky-position distributions of
outflow material must reflect the combined distributions of
density in the ambient medium and forces or momentum fluxes that drive the
molecular flow.  Hence, although mass and momentum maps and line profiles
from outflows are not directly invertible to yield the full
outflow density distribution $\rho({\bf v},{\bf x})$, they can be used to help 
discriminate among different proposals for the nature of the primary wind and
outflow acceleration mechanism (e.g. \cite{mass92}, \cite{che95}, 
\cite{nag97}).  
Models in
which the outflow consists of a shell of ambient matter swept up by a 
wide-angle
radial primary wind in a momentum-conserving fashion appear able to account 
for many observed outflow properties, provided appropriate stratification
exists
in both wind and ambient media (\cite{shu91}, \cite{li96c}).  In alternative 
models, the
primary wind is assumed to be jetlike (axial flow velocities) and the
molecular outflow is the manifestation of a bow shock in the ambient medium
(e.g. \cite{blo90}, \cite{mass93}, \cite{rag93}, \cite{sto93},\cite{che94}).
In this case the component of the outflow momentum transverse to the jet is 
driven by pressure gradient forces in the working surfaces at the jet head
and within the jet (when variability leads to internal shocks),
and the agreement with observed outflow properties depends very sensitively
on the cooling rate of shocked gas.

The present uncertainty about outflow acceleration mechanisms highlights (and
derives from) a more basic uncertainty regarding the origin and nature of the 
primary wind which is ultimately responsible for both the observed jets and
outflows.
Because observed velocities of optical jets are comparable to stellar escape
speeds, the jet material almost certainly originates near the star.
Jets appear quite collimated in density down to distances within a few tens of
AU of the source (\cite{ray96}).  
The observed optical jets are likely only the ionized, strongly-emitting inner
portions of a broader, neutral wind with lower density and lower 
outflow speed than the jet (for observational evidence, see e.~g.
\cite{hea96}, \cite{har97} and references therein).  The presence of separate
high and low velocity blueshifted components in forbidden line profiles
suggests that these winds may originate over a range of radii in the
circumstellar disks (\cite{har95}).
An open question, however, is whether the apparent collimation of optical
jets is due to cylindrical {\it density} stratification of a 
primarily radial MHD wind which originates near the star 
(\cite{shu95}, \cite{li96b}), or whether 
there is an MHD wind originating in an extended region of the disk 
(out to $\sim 100$ AU) which is itself well-collimated in {\it both}
velocity and
density, and which helps to collimate the observed jet and drive the 
larger-scale outflow (e.g. \cite{pud86}, \cite{kon93}).
If observations end up demanding the latter, many uncertainties
remain in models for MHD disk winds, especially in relation to the
requirements for producing a collimated (but not recollimated) flow without
singularities or other unphysical behavior.  
In the remainder of this Introduction, we describe current ideas about
MHD winds driven from rapidly-rotating accreting systems (``magnetocentrifugal
winds''), and discuss some of the difficulties in previous models of disk
winds.  This motivates the present work, which develops models for 
magnetocentrifugal winds that are constrained 
to become fully collimated at large distances from the source, and 
describes the general properties of such flows.

\subsection{Magnetocentrifugal Disk Winds}

The basic physics of steady magnetocentrifugal winds has been outlined in
numerous sources; for a recent pedagogical review, see \cite{spr96}.
\cite{hey89} and \cite{hey96} have used analytic arguments to predict how
asymptotic streamline collimation should develop in winds 
with varying properties 
(an analogous treatment for the relativistic case is given by \cite{chi91}).
The mixed hyperbolic and elliptic nature of the general
governing equations (\cite{hei78}), however, leads to technical
difficulties in formulating and finding exact solutions for the
steady-state problem in terms of appropriate boundary conditions (see
e.g. \cite{bog94}), due to the presence of critical surfaces --
projected to curves in the poloidal plane -- within the flow.
Nevertheless, steady-state, fully two-dimensional solutions have been
obtained for the specific cases of a ``split-monopole'' poloidal
magnetic field interior boundary condition (\cite{sak85},
\cite{sak87}), and for the case of an ``X-point'' poloidal magnetic
field interior boundary condition -- strongly pinched magnetic fields fanning
out of the disk near the star (\cite{shu94b}, \cite{naj94}, 
\cite{shu95}).  

To circumvent the difficulties associated with free critical surfaces, 
an approach that was earlier adopted by \cite{bla82} (hereafter BP) 
is to look for families of solutions with certain pre-determined symmetries.  
In BP, the fundamental assumption is that 
all velocities obey the same $v\propto r^{-1/2}$ scaling with spherical radius 
as the Kepler velocity, and that the density obeys $\rho\propto r^{-3/2}$
(the scaling associated with a mass-conservative spherical wind radially
flowing in a Kepler potential).  The magnetic field
components then must behave as $B\propto r^{-5/4}$ in order to have the 
corresponding \alf\ speed $\va^2=B^2/(4\pi\rho)\propto r^{-1}$.
By assuming these scalings for the radial dependence of all quantities, the
governing PDEs of the MHD wind problem are converted to ODEs in angular
coordinate $\theta$ (or, as many workers have framed the problem, in the
scaled height above the midplane).  The assumed similarity scaling results
in any critical surface coinciding with a radial line ($\theta=const.$).
\cite{bla82} obtained a family of solutions which successfully navigated the
\alf\ transition, directly demonstrating that cold MHD winds from rotating 
disks can reach large velocities and collimate their streamlines
toward the poles.

The BP $r$-self-similar solutions have subsequently been
extended and generalized by other workers, including 
relativistic solutions by \cite{li92} and \cite{con94b}, and 
nonrelativistic solutions with different basic scalings by \cite{con94} 
(hereafter CL) and
\cite{fer96}.  In addition, \cite{sau94} and \cite{tru96} have obtained MHD 
wind solutions with different basic symmetry (latitudinal, rather than
radial, self-similarity) via alternative choices for separation of
variables, for application to flows from rapidly-rotating stars rather
than accretion disks.  

Eventually, a theory of magnetocentrifugally driven winds should be able
to treat time-dependent, non-axisymmetric configurations with a fully
self-consistent connection to the disk or star where the flow
originates, and to the surrounding medium.  Time-dependent,
axisymmetric simulations of outflows from accretion disks
have begun to be pursued by, e.g.
\cite{uch85}, \cite{shi86}, and \cite{sto94} (who include the internal
disk dynamics), and by \cite{ust95}, \cite{kol95}, and \cite{ouy97}
(who treat the disk as a boundary condition for the wind).  The simulations
produced so far verify that well-collimated MHD outflows (sometimes 
with intriguing time variability) can be generated
by differentially-rotating disks threaded by a mean magnetic field, and that 
rapid disk accretion can occur as a result of angular momentum removal
by the wind.  However, the major technical 
effort required to produce these simulations (as well as the steady 
fully-2D solutions cited above) makes it difficult to explore parameter space
extensively.  Thus, it remains quite useful to study the properties of
magnetocentrifugally driven outflows with models that impose a 
symmetry in advance to reduce the problem to coupled ODEs.  
The $r$-self-similar ansatz may be particularly 
appropriate for modeling outflows from accretion disks with a large
dynamic range of radii, in which the flow may approach a scale-free 
solution well away from the boundaries.  Indeed, the simulations of
\cite{ouy97} demonstrate that steady, $r$-self-similar winds may
naturally develop when a mass flux is driven from the surface of a magnetized,
rotating disk.

\subsection{Wind-Driven Accretion}

In general, accretion in a disk may be driven in part by local stresses 
(acting on radial scales comparable to the disk thickness, and often 
parameterized by an ``$\alpha$'' viscosity, e.g. \cite{pri81}), and in part by 
larger-scale forces such as those associated with global spiral density 
waves (e.g. \cite{spr87}) or an MHD disk wind (e.g. BP, 
\cite{pud86}).  An interesting special case of an accretion disk/wind system
is the extreme one where the disk wind from a given annulus removes
the angular momentum and energy needed for what is left of the 
disk to accrete to the next annulus closer to the star. 
If we assume a steady state for both wind and disk, 
the conservation equations yield
a differential mass transfer rate $|d\dot M_W/dR|=|d\dot M_D/dR|$
at radius $R$, with 
\begin{equation}
{d\ln\dot M_W\over d \ln R}\left({J\over \Omega R^2} -1\right) = {1\over 2},
\end{equation}
in terms of the specific angular momentum $J$ in the wind and the local 
Kepler rotation rate of the disk $\Omega$ (this expression assumes a thin disk
in Kepler rotation).  For self-similar solutions where
density $\rho\propto r^{-q}$ in both the wind and disk, this would imply
that the wind requires a ratio of the cylindrical \alf\ radius $\ra$ to the 
streamline footpoint radius $R_0$ of 
\begin{equation}\label{STEADYREQ}
{\ra\over \ro}=\left({4-2q\over 3-2q}  \right)^{1/2}
\end{equation}
because $J=\Omega \ra^2$
(see \S2.1).  Solutions which lose very little mass in the outer disk
($q$ nearly $3/2$) must carry large specific angular momentum, and
conversely, solutions with massive winds from their outer disks (small
$q$) must have relatively low specific angular momenta in their winds
if they are to be consistent with a steady state inflow/outflow.   

Even if the \alf\ radius does not satisfy equation (\ref{STEADYREQ}), a
self-similar MHD disk wind will drive inflow in the disk with the
disk accretion rate $\dot M_D(R)$ a power-law in $R$.  For these more
general cases, the surface density at any point in the disk would
either increase or decrease in time depending on whether $\ra/\ro$ is
greater or less than the value in equation (\ref{STEADYREQ}); the local mass
deposition/removal timescale would be greater than the local accretion
timescale by a factor $2\left[3 - 2q - 1/((\ra/\ro)^2-1)
\right]^{-1}$.

The connection between mass and angular momentum loss in the wind and 
accretion in the disk was explored by \cite{kon89} and \cite{war93}, 
using local models for the disk and connecting to the BP wind 
solutions;  they showed that the field geometry required for a disk wind
can be self-consistently provided by a diffusive (ambipolar or Ohmic) disk.
The work of \cite{li95} and \cite{fer95} incorporating additional dynamics
supported these conclusions.  Most recently,
\cite{fer96} has obtained steady, $r$-self-similar combined inflow/outflow
solutions for the case of ordinary resistivity in the disk region,
while \cite{li96a} has obtained steady inflow/outflow solutions assuming
ambipolar diffusion in the disk (although neither set of solutions
treats the wind far from the disk in a complete fashion; see below).
In both of the last two cases, the solutions have $q>1.4$,
corresponding to a relatively small fraction of the disk mass being
lost to a wind.

\subsection{Critical Points and Wind Asymptotics}\label{WINDASY}

An outflow that accelerates from low velocity near the surface of an
accretion disk to high velocities at a distance (permitting escape
from the gravitational potential) may pass through several points
where the flow changes physical character. Such transitions occur when
the flow speed surpasses the speed of an allowed wave mode (sound
waves for unmagnetized or MHD waves for magnetized winds), and are
manifest by apparent singularities in the equations governing the
flow.  Accounts of the nature of critical points in general and
restricted MHD winds are given, e.~g., by \cite{hei78}, \cite{spr96},
\cite{tsi96}; we summarize some of the main points here.

An important issue is whether, and where, the equations
pass from being elliptic to hyperbolic at large distances from the
source.  Fully two-dimensional (axisymmetric) MHD flows become
hyperbolic when the poloidal speed $\vp$ exceeds the
fast-mode MHD wave speed in the the poloidal direction
$\vf=(1/2)\left\{ \cs^2 + \va^2 + 
\left[(\cs^2+\va^2)^2 - 4\cs^2\vap^2 \right]^{1/2}\right\}$ 
(where $\cs$ is the sound speed, $\va$ is the \alf\ speed, and
$\vap$ is the \alf\ speed associated with the poloidal magnetic field 
component $\Bp$). For a cold flow where $\vf\rightarrow \va$, 
the fast-mode Mach number for the poloidal flow is 
$\mf\equiv \vp/\va$;  the 2D equations 
become hyperbolic when $\mf>1$.  The full 2D equations also have an
apparent singularity at the \alf\ critical point where $\vp/\vap\equiv\ma =1$;
here the fluid speed in the poloidal direction equals the wave speed
of the \alf\ mode propagating in the poloidal direction.

As first pointed out by BP and recently reexamined in detail by
\cite{tsi96}, the restriction to an $r$-self-similar model alters the
nature of the critical points.  With the assumption of
$r$-self-similarity, the poloidal-plane PDE for cross-field force
balance is converted to a second-order ODE in $\theta$.  Critical
points in the equations now occur when the coefficient of the
highest-order $\theta$-derivative in the cross-field equation passes
through zero.  The modified fast-MHD and \alf\ points occur in the
$r$-similar MHD equations, respectively, if $|v_\theta|/v_{f,\
\theta}= 1$ (where $v_{f,\ \theta}$ is obtained by replacing $v_{\rm
A,\theta}$ for $\vap$ in the above expression for $\vf$), and if
$|v_\theta|/v_{\rm A, \theta} =1$.  For a cold flow, $\vf=v_{f,\
\theta}=\va$; the fast-MHD wave speed is the same for all propagation
directions. For an $r$-similar flow the modified fast-MHD critical
point would occur at $\mf(|v_\theta|/\vp)=|v_\theta|/\va=1$, beyond
the point $\mf=\vp/\va=1$ where the fast-mode transition occurs (there
are now no singularities in the equations at the location where
$\mf=1$).  Because $|v_\theta|/v_{\rm A, \theta} = |\vp|/\vap$, the
\alf\ point in the $r$-similar flow still occurs at $\ma=1$.

Physically, the modification of critical points happens because
$r$-self-similar model flows remain {\it a priori} in effective causal
contact in the $\hat r$ direction (as well as in the $\hat \varphi$
direction for assumed axisymmetric flows) for arbitrary Mach number.
Thus critical transitions respect just the projection of the velocity
in the direction of the spatial degree of freedom, $\hat \theta$,
relative to the propagation speeds of the various wave families.  In
particular, effective ``ellipticity'' in the $r$-self-similar reduced
equations is maintained as long as the $\hat \theta$ component of the
velocity is smaller than the fast-MHD wave propagation speed in the
$\hat \theta$ direction, $v_{f, \theta}$ ($=\va$ for a cold flow).
Geometrically, the condition is that the $r$-similar flow becomes
effectively ``hyperbolic'' when the projection of the ``minus
characteristic'' in the $\hat \theta$ direction becomes negative
(cf. \cite{tsi96}; \cite{con95}).  The corresponding 2D equations
remain {\it formally} hyperbolic (in the sense that characteristic
curves are defined) in the whole region of the poloidal plane where
$\mf>1$. Since, however, only the $\hat\theta$ direction is a spatial
degree of freedom, only the projections of the characteristics in the
$\hat\theta$ direction are relevant for the propagation of information
from the boundaries into the body of the solution.  Thus only if the
flow were to pass through and remain at $|v_\theta|/\va>1$ could we
regard the information about the nature of boundary conditions at the
pole as irrelevant to the solution.  Even if such a solution were
found, its relevance to real winds would be questionable because the
propagation of information from the boundaries in the radial direction
is ignored a priori.

More generally, full specification of the steady magnetocentrifugal wind
problem requires both the flow equations and a statement of the
desired boundary conditions.  Whether or not the flow becomes hyperbolic,
the interface with the ambient medium inside and outside the first and
last streamlines of the outflow must, at least in principle, govern
the latitudinal extent of a steady state wind (e.~g. \cite{shu95}).  
Since, however, the $r$-self-similar model
is infinite radially -- with infinite flux -- there is no proper
interior or exterior to the wind. The only solution boundaries for a
cold, $r$-similar wind lie along $\theta=0$ and $\theta=\pi/2$.  The
boundary at $\theta=\pi/2$ corresponds to the disk surface, where the
solution should be able to match to a subsonic flow; several authors
have discussed the additional constraints this imposes (see \S 1.3).

The boundary at $\theta=0$ corresponds to the asymptotic limit of the flow
at a large distance from the source.  Even if an $r$-self-similar solution
covers all angles, hence formally filling space and leaving no room
for an ambient medium to match, a solution should be ``physically
reasonable'' in the sense that if spatially truncated, the solution
could be embedded within a larger (non-self-similar) solution without
any very particular requirements.  Thus, we would like to obtain
self-similar solutions where the $\theta\rightarrow 0$ 
asymptotics could sensibly be matched
to an ambient medium with generic properties.  Such a matching
would select among possible input parameters those which yield
acceptable asymptotic solutions. 

Previous $r$-similar MHD wind solutions containing both toroidal and
poloidal fields -- for a variety of scaling parameters -- have not
explicitly implemented matching to boundary conditions at the pole.
In some work, numerically-integrated solutions are halted when they
become singular near the modified fast-MHD point at 
$|v_\theta|/\va=1$, since no regularity condition is applied 
(e.g. \cite{fer96}).  Other
solutions are numerically integrated away from the equator and halted
at an arbitrary point near the pole, with $|v_\theta|/\va<1$ throughout the
computed region (e.~g. BP and CL).  
A difficulty with these solutions is that many implicitly 
require special boundary conditions.  Namely, a large class of
solutions (with $q>1$) generically recollimates (i.e. the cylindrical
radius $R$ of a streamline reaches a maximum and turns around), and
another large class of solutions (with $q<1$) generically shows radial
oscillations of the streamlines (CL; see also \cite{sau94}).
\cite{con94} found a third class of solutions (with $q>1$)
which does not show recollimation within the numerically-integrated
regime, but they do not explicitly implement a $\theta\rightarrow
0$ asymptotic boundary condition.

On the other hand, \cite{con95} has found a related solution with
purely toroidal fields (where the flow does not accelerate from zero
speed at the disk but instead requires a large initial poloidal
injection speed, and since $\vap\equiv 0$ there is never an \alf\
transition) with the input parameters tuned such that the solution
makes a transition through the modified fast-MHD point
$|v_\theta|/\va=1$.  While solutions containing both toroidal and
poloidal fields can in principle make a transition through
$|v_\theta|/\va=1$, examples of such solutions have not yet been
obtained due to the additional numerical effort demanded.  The added
requirement of making the $|v_\theta|/\va=1$ transition would lower
the number of free parameters in the solution by one, so that, for
example, the two-parameter family of cold wind solutions of BP would become 
a one-parameter family.

In the present work, we set out find MHD wind solutions which {\it do}
match to a specific asymptotic prescription.  Inspired by observations
which show strong collimation in both $\rho$ and $\bf v$, we seek
self-similar solutions which have cylindrical asymptotics.  Thus, while our
basic set of dynamical equations is equivalent to the cold wind equations
of CL or \cite{fer96} (although mathematically represented quite differently), 
we supplement them with an
additional boundary condition at the pole ($\theta=0$).  The
solutions we find are not global, in the sense that they do not
directly connect onto the interstellar medium, or to the medium
interior to the outflow (e.g. hot plasma, axial fields, or a
mass-carrying MHD wind originating from the central object/inner
disk).  Nevertheless, the uniform cylindrical asymptotic behavior of
these solutions allows a conceptual matching onto simple boundary
conditions of constant (high) pressure interior, and (low) pressure
exterior, to the region of the outflow.  The levels of the ambient
pressure would then select among the possible solutions by matching
pressures at the boundaries.  Although the condition of exactly cylindrical
asymptotic collimation is probably more extreme than most real jet/outflows,
the properties of these solutions makes clear the stark contrast
in behavior between collimated and uncollimated magnetized flows.

\subsection{Plan of Paper}

In \S2, we first (\S2.1) state the governing equations for a steady,
axisymmetric, MHD flow, then (\S2.2) present an exact analytic
solution for self-similar, differentially-rotating cylindrical flows
which represent the chosen asymptotic boundary conditions far from the disk,
and finally (\S2.3) detail the $r$-similar wind {\it ansatz} 
and the resulting reduced forms of the wind equations.  \S3 presents
solutions for disk winds that satisfy the self-similar equations with  
cylindrical asymptotics, and \S4 discusses the properties of these solutions
and compares to previous models and observations.  
Various mathematical details are described in appendices \S \S A.1-A.4



\section{Magnetocentrifugal wind equations}

\subsection{Steady, axisymmetric ideal MHD equations}\label{IDEQS}

Consider a steady, axisymmetric magnetized flow with negligible resistivity.
From axisymmetry, the poloidal magnetic field can be expressed in terms of
a flux function $\Phi$ as
\begin{equation}\label{BPDEF}
\bB_p = {-\phihat\times\nabla\Phi\over R},
\end{equation}
where $R=r\sin\theta$ is the cylindrical radius.  From the $\zhat$
component of the induction equation $\nabla\times(\bv\times\bB)=0$, it can be 
shown that $\Phi$ is
conserved on streamlines, $\bv_p\cdot \nabla\Phi=0$, and that
\begin{equation}\label{BETADEF}
\bB_p=\beta\rho\bv_p
\end{equation}
for some $\beta$ -- i.e. the poloidal streamlines and fieldlines are 
parallel.  Combining the continuity
equation $\nabla\cdot(\rho \bv)=0$ 
and $\nabla\cdot\bB=0$ shows that $\beta$, which represents
the ratio of magnetic field to mass flux, is also conserved on
streamlines: $\beta=\beta(\Phi)$ and $\bv_p\cdot\nabla \beta=0$.  Defining 
\begin{equation}\label{OMEGADEF}
\Omega\equiv [\vphi -\bphi/(\beta\rho)]/R
\end{equation}
and 
\begin{equation}\label{UDEF}
\bu\equiv\bv-\Omega R\phihat,
\end{equation} 
we obtain $\bB=\beta\rho\bu$, i.e. the flow in a frame
rotating at $\Omega$ is frozen to the local field line.  Using
$\bv\times\bB= \Omega\nabla \Phi$ and the induction equation, one can
show that $\nabla \Omega\times\nabla \Phi =0$, so that $\Omega$ is
constant along field lines.  Thus $\Omega(\Phi)$ is the rotation rate
of a given field line, and on the field line labeled by $\Phi$ in the frame 
rotating at $\Omega$, the flow and field are parallel, $\bu\parallel\bB$.
Each field line, of course, can have a different rotation rate $\Omega(\Phi)$. 
From equation (\ref{UDEF}), $\bu_p =\bv_p$;  we use the symbols interchangably.

The toroidal component of the momentum equation yields an additional conserved
quantity along field lines, the specific angular momentum $J$ associated with
the combined matter flow and Poynting flux
\begin{equation}
J=J(\Phi)=R(\vphi - {\beta\bphi\over 4\pi}).
\end{equation}
Using the definition (\ref{UDEF}), we write 
\begin{equation}\label{JDEF}
J= \Omega R^2 + R \uphi(1- \ams)\equiv \Omega \ra^2,
\end{equation}
where 
we use the definition for the \alf\ Mach number $\ma$ of the flow
\begin{equation}\label{MADEF}
M_{\rm A}^2 \equiv {\vp^2\over v_{\rm A, p}}={\vp^2\over  \Bp^2/(4\pi\rho)}
={4\pi\over \beta^2 \rho}. 
\end{equation}
Note that $\vp/v_{\rm A, p}=u_\varphi/v_{\rm A, \varphi}=|\bu|/\va=\ma$.
The \alf\ radius $R=\ra$ is the point where a flow makes a transition from 
$\ma<1$ near the disk to $\ma>1$ in the wind.

Taking the component of the momentum equation in the direction of
$\bB$ yields the Bernoulli equation, which states that
\begin{equation}\label{BERNEQ}
\cale\equiv {1\over 2}|\bu|^2 +{\gamma\over \gamma -1} {P\over \rho} 
+V_{\rm g}-{1\over 2} \Omega ^2 R^2
\end{equation}
is constant on field lines, $\cale=\cale(\Phi)$.  Here, 
we have assumed an ideal gas with ratio of specific heats $\gamma$,
and throughout this work we shall assume the gravitational
potential $V_{\rm g}=GM/r$ of a central point mass.  
In the absence of heating and cooling, the final conserved 
quantity on field lines is the specific entropy, such that
$K\equiv P\rho^{-\gamma}=K(\Phi)$.

The final dynamical equation is obtained by taking the component of the
momentum equation along $\nabla \Phi$.  The resulting 
Grad-Shafranov equation describes
force balance in the direction perpendicular to the poloidal field
$\bB_p$,  and can be written
\begin{equation}\label{GSEQ}
\nabla\cdot \left[(1-\ma^2){\beta \nabla \Phi\over 4\pi R^2} \right]=
{\bphi\over R} {d J\over d\Phi} - \beta\rho{d\cale\over d\Phi}
+ {|\bB|^2\over 4\pi}{d\beta\over d\Phi} - R\beta\rho\vphi{d\Omega\over d\Phi}
+ {\beta \rho^\gamma\over \gamma-1} {d K\over d\Phi}.
\end{equation}

In this work, we will find it convenient to express the various flow and 
field variables in terms of $\Phi$ and the conserved quantities $\beta$,
$J$, $\Omega$, $\cale$, and $K$.
The toroidal speed in the rotating frame is obtained from equation (\ref{JDEF})
as
\begin{equation}\label{UPHIEQ}
\uphi={J/R-\Omega R\over 1 - \ams};
\end{equation}
the inertial-frame toroidal speed is 
\begin{equation}
\vphi={J/R-\ams \Omega R\over 1 - \ams}.
\end{equation}
The magnitude
of the poloidal speed can be obtained from equation (\ref{BERNEQ}) as
\begin{equation}\label{UPEQ}
|\bu_p|^2 = 2\cale - 2 V_{\rm g}+ \Omega^2 R^2 - 
\left({J/R-\Omega R\over 1 - \ams}\right)^2 - 
2 {\gamma\over \gamma -1} {P\over \rho} .
\end{equation}
An alternative expression for the poloidal speed comes from the definition 
(\ref{BPDEF}) together with $\bu_p=\bB_p/\beta\rho$.  Equating the two
expressions yields an equation for $\rho$.  For the case of a cold flow with
$P=0$, $\rho$ is the solution of a quartic equation in terms of $\Phi$,
$\nabla \Phi$, $\beta$, $J$, $\Omega$, $\cale$ and $r,\  \theta$.

\subsection{Self-similar magnetized cylindrical flows}\label{CYLEQS}

In this section, we consider the steady, ideal, axisymmetric MHD
equations without gravity, and find a class of solutions representing
rotating, cylindrically self-similar flows.  In these solutions, the
poloidal velocity is everywhere axial ($\bv_p \parallel \hat z$), all
speeds scale with cylindrical radius $R=r\sin\theta$ as 
$v\propto R^{-1/2}$, the
density scales as $\rho\propto R^{-q}$, and all components of the
magnetic field scale as $B\propto R^{-(1+q)/2}$.  All flow variables
are independent of $z$.  The cylindrical solutions have identical
similarity scaling in $\sin\theta$ to their similarity scaling in $r$.
These cylindrical solutions define the {\it asymptotic} angular
behavior for the radially self-similar disk outflows considered in
this work, which satisfy the general $r$-self-similar equations to be 
presented in \S\ref{SSEQS} (i.~e. the same radial scaling as the asymptotic 
state but arbitrary scaling in $\theta$).  For simplicity,
we specialize here to cold flows ($P=0$), but the more general
expressions are easily derived.

Starting from the above {\it ansatz} for scalings, any solution must have 
$\Phi\propto R^{3-q\over 2}$, $\Omega\propto R^{-3/2}$,
$\beta\propto R^{q/2}$, $J\propto R^{1/2}$, and $\cale\propto R^{-1}$.
The \alf\ Mach number $\ma$ is uniform throughout the flow.  In order to 
normalize these power-law solutions, we choose a
fiducial field line $\Phi=\Phi_1$ lying along $R=\rl$, and rotating
at a rate $\Omega_1=\Omega(\Phi_1)$.  We define 
$j\equiv J/\Omega R^2$, $e\equiv -\cale/(\Omega R)^2$, and 
$m\equiv \ams={\beta^2\rho\over 4\pi}$,
all of which quantities are independent of $R$.
Then, normalizing all speeds by
$\Omega R=\Omega_1 R_1 (R/R_1)^{-1/2}$, we have inertial-frame toroidal 
velocity
\begin{equation}\label{VPHICYL}
{v_\varphi\over \Omega R} ={j-m\over 1-m}, 
\end{equation}
rotating-frame toroidal velocity 
\begin{equation}\label{UPHICYL}
{u_\varphi \over \Omega R}= {j-1\over 1-m}, 
\end{equation}
poloidal speed $v_p=v_z$
\begin{equation}\label{VPCYL}
{|v_p|\over \Omega R}= \left[ 1-2e - {(j-1)^2\over(1-m)^2}\right]^{1/2},
\end{equation}
and toroidal and  poloidal \alf\ speeds
\begin{equation}\label{VAPHICYL}
{\vaphi} = m^{1/2}u_\varphi
\end{equation}
and 
\begin{equation}\label{VAPCYL}
{\vap} = m^{1/2} \vp.
\end{equation}

For the cold, zero-gravity solutions, the Bernoulli equation (\ref{BERNEQ}) 
with the cylindrical {\it ansatz} reduces to
\begin{equation}
{|\bu|^2\over (\Omega R)^2} = 1-2e,
\end{equation}
so that $v_A^2/(\Omega R)^2=m(1-2e)$. 
The only remaining constraint on the flow is the Grad-Shafranov equation 
(\ref{GSEQ}), 
which for our present assumption of a cylindrically-stratified flow is just
the $\hat R$ component of the momentum equation,
\begin{equation}
0={v_\varphi^2\over R} 
-{1\over 8\pi \rho R^2}{d (R^2 B_\varphi^2)\over d R} 
-{1\over 8\pi \rho }{d (B_p^2)\over d R},
\end{equation}
where we have dropped the thermal pressure term assuming a cold flow (i.e. 
$c_{\rm s}$ 
small compared to the flow and \alf\ speeds).  Now employing the self-similar
scaling for the magnetic field $|\bB|\propto R^{-(1+q)/2}$, this reduces to
\begin{equation}\label{CYLGS}
0=v_\varphi^2 +{q-1\over 2}v_{\rm A, \varphi}^2  + {1+ q\over 2} v_{\rm A, p}^2.
\end{equation}
Substituting in equation (\ref{CYLGS}) for 
$v_\varphi$, 
$\vaphi$, and $\va$ in terms of $e$, $j$, and $m$, we arrive at the 
quadratic equation
\begin{equation}\label{MAQUAD}
0= m^2(1-2e) {1+q\over 2} + m\left[{1-q\over 2} + (1+q)e \right] -j^2.
\end{equation}
The physical solution has
\begin{equation}\label{MACYL}
\ma^2\equiv m^{-1}=(2 j^2)^{-1} 
\left\{ {1-q\over 2}  + (1+q)e + 
\left[\left({1-q\over 2} + (1+q)e\right)^2 +2j^2(1+q)(1-2e) \right]^{1/2}\right\}.
\end{equation}
Thus, given $q$ and the values of the angular momentum and Bernoulli 
parameters 
$j$ and $e$, we obtain cylindrical solutions by substituting equation
(\ref{MACYL}) in equations (\ref{VPHICYL})-(\ref{VAPCYL}).  
The fast-mode Mach number $\mf\equiv |\bv_p|/\va$ is also constant 
throughout the cylindrical flow, and is written
\begin{equation}\label{MFCYL}
\mf^2=\ma^2 \left[1- {1\over(1-2e)} \left({1-j\over 1-m}\right)^2   \right].
\end{equation}  
When $1-q>>j,\ e$, an approximate solution to equations 
(\ref{VPHICYL}-\ref{MFCYL}) is 
$\vphi/(\Omega R)\approx j$, $\vp/(\Omega R)\approx (2j-2e)^{1/2}$, 
$\vaphi/(\Omega R)\approx j[2/(1-q)]^{1/2}$, 
$\vap/(\Omega R)\approx 2j(j-e)^{1/2}(1-q)^{-1/2}$, 
$\ma\approx j^{-1}[(1-q)/2]^{1/2}$, and 
$\mf\approx j^{-1}(1-q)^{1/2}(j-e)^{1/2}$.

These self-similar cylindrical solutions only exist for a limited range
of parameters.  In particular, the original Blandford-Payne scaling
$\rho\propto r^{-3/2}$, $B\propto r^{-5/4}$ is not compatible with the
self-similar cylindrical asymptotic solutions described here.  In
fact, equation (\ref{CYLGS}) can only be satisfied only for $q<1$,
i.e. when the density and magnetic field profiles are less steep than
$R^{-1}$ (the same criterion also holds when thermal pressure is included).
Physically, this is true because the tension associated with the
toroidal field is the only inward force that can oppose the outward
centrifugal force and outward (for $q>-1$) force associated with the
magnetic (and thermal) pressure gradient.  Only for $q<1$ is the hoop
stress large enough to enforce cylindrical collimation, and the closer
$q$ is to 1, the larger the collimation radius relative to the launch
point.  Notice that it is only the inclusion of a toroidal
field $\vaphi\ne0$ that permits cylindrical solutions for flows in
which the magnetic pressure decreases outwards.
In cases where the magnetic pressure (i.e. the fluid energy density) 
increases outward ($q<-1$), both the poloidal and toroidal components of the 
field apply stresses that oppose the centrifugal force in equation 
(\ref{CYLGS}), and collimation occurs at relatively small radii.  However,
we do not consider cases with $q<-1$ likely to be appropriate models for winds
from extended accretion disks, so we do not consider them further herein.

The axial current carried by the self-similar cylindrical flow scales as
$R^{1-q\over 2}$.  Therefore, for $q<1$ the current increases with the 
radial scale of the flow, and there is no singularity at the axis.  
The total mass, momentum, and energy per unit 
time carried by the flow within $R$ scale as $\dot M\propto R^{3/2-q}$, 
$\dot P \propto R^{1-q}$, and $\dot E \propto R^{1/2-q}$.  Thus the cylindrical
solutions with $q<1$ have mass and momentum flows dominated by the outer
regions, and energy flow dominated by the interior (exterior) for $q>1/2$ 
($q<1/2$).  

\begin{figure}
\plotone{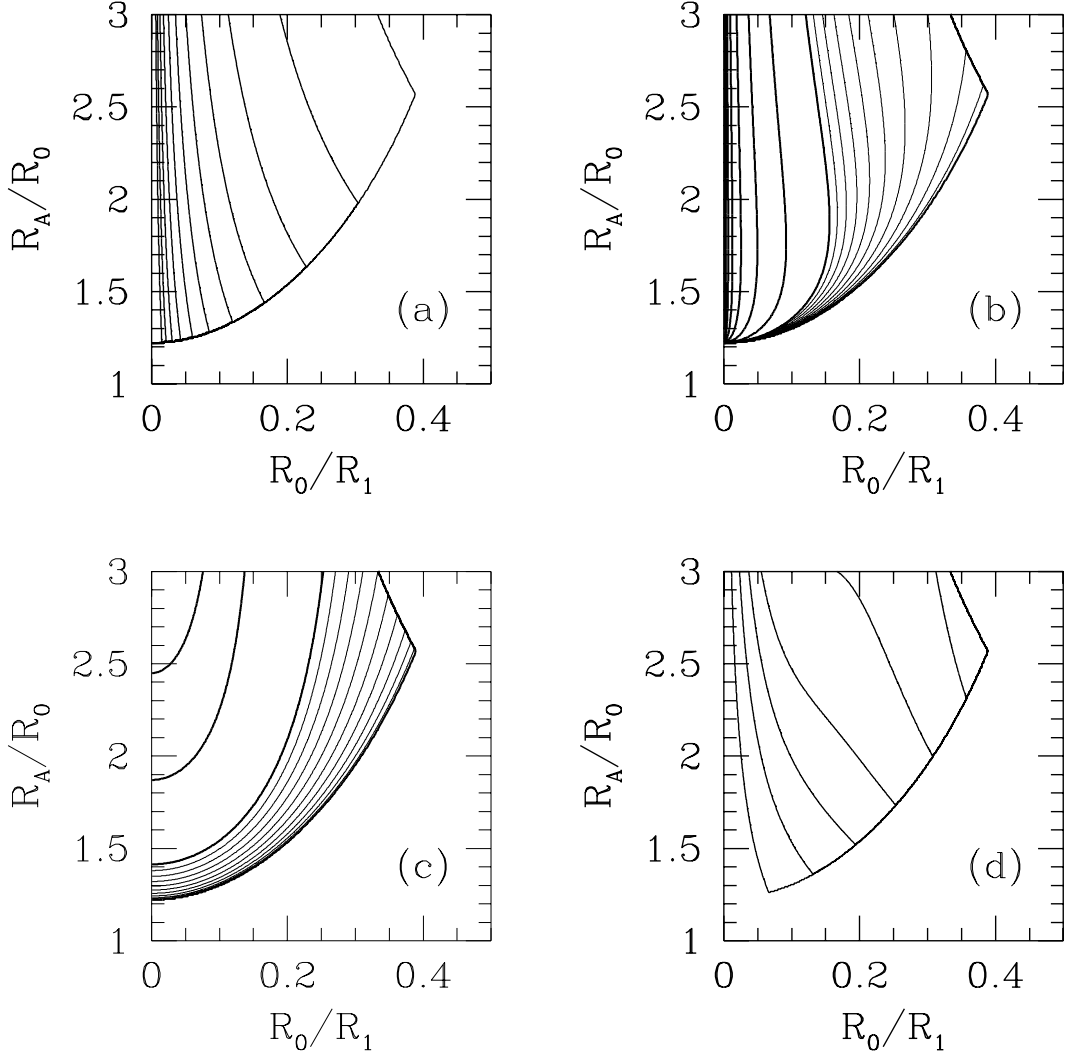}
\caption{
Parametric dependence on $\ro/\rl$ and $\ra/\ro$ 
of the asymptotic self-similar cylindrical solution family with $q=0.75$.  
For a given cylindrical solution with field line rotation rate $\Omega$, 
specific angular momentum $J$ and Bernoulli constant ${\cal E}$ on the
streamline through $\rl$, $\ro$ and $\ra$ represent the cylindrical radii
of the footpoint and \alf\ point a disk outflow would need in order to
have the same $\Omega=\sqrt{GM/\ro^3}$, 
$J=\Omega \ra^2$, and  ${\cal E}= -(3/2)(\Omega \ro)^2$.
(a) Alfv\'en Mach number $\ma=\vp/\vap$.  Contours show
$\ma=1,2,4,8,...$, from right to left.  
(b) Fast MHD Mach number $\mf=\vp/\va$.
Heavy contours show $\mf=0,1,2,4,8,...$ from right to left; light contours
$\mf=0.1,0.2,0.3,...$.  (c) Poloidal speed relative to Kepler speed on
the disk.  Heavy contours show $\vp/\Omega\ro=0,1,2,3$, from right to
left; light contours show $\vp/\Omega\ro=0.1,0.2,0.3,...$.  (d)
Toroidal speed relative to Kepler speed on the disk.  Contours show
$\vphi/\Omega\ro=0.1,0.2,0.3,..,0.6$ from left to right.
}
\end{figure}

A further constraint on the parameter space for which self-similar cylindrical
solutions exist is that the solution of equation (\ref{VPCYL}) be real.  The
boundary of parameter space is found by equating $|\bv_p|=0$ and using
equation (\ref{MACYL}); solutions exist 
for
\begin{equation}\label{JMIN}
j\geq{2 e \over 1 + (1-2e)^{1/2}}\left[1+ {4e\over(1-q)} 
{1\over \left(1 + (1-2e)^{1/2}\right)(1-2e)^{1/2}}\right].
\end{equation}  All solutions must have $e<1/2$.

If we think of a given cylindrical solution as the asymptotic limit of 
an outflow from an accretion disk, we can express the parameters $j$ and 
$e$ in terms of important physical scales in the flow.  
From equation (\ref{JDEF}), $j=(\ra/\rl)^2$, where $\ra$
is the \alf\ radius for the streamline that asymptotes to $\rl$.  From 
equation (\ref{BERNEQ}), the Bernoulli parameter is equal to its
value where the flow leaves the disk.  If the same cold flow streamline 
originates in a Keplerian disk at launch radius $\ro$, 
$\bu=0$ on the disk surface so $\cale= -(3/2)(\Omega \ro)^2$ . 
Using $\Omega^2=G M/\ro^3$ for the fiducial streamline, 
$e=(3/2)(\ro/\rl)^2$.  The flow speeds $\vphi$ and $v_z$ 
relative to the Kepler speed $\Omega \ro$ at the footpoint of any streamline 
are found by multiplying the right-hand-sides of equations (\ref{VPHICYL})
and (\ref{VPCYL}) by $\rl/\ro$.  From equation (\ref{VPCYL}), the ratio of
the final speed to the Kepler speed at the launch point is bounded by 
\begin{equation}\label{VPLIMIT}
{\vp\over \Omega\ro}<\left[2\left({\ra\over\ro}\right)^2 -3\right]^{1/2}.
\end{equation}

Figure 1 shows the dependence of $\ma$, $\mf$, $\vphi$, and $v_p$ on
the parameters $\ro/\rl\equiv(2e/3)^{1/2}$ and 
$\ra/\ro\equiv[3j/(2e)]^{1/2}$ for an example of the cylindrical
solution family 
with $q=0.75$.  At other values of $q$, the behavior is similar,
with the right-hand ``corner'' of the solution moving toward smaller 
$\ro/\rl$ and larger $\ra/\ro$ as $q$ increases.

From equation (\ref{JMIN}), $j>e$ so a cold flow originating in a 
Kepler-rotating disk must have $(\ra/\ro)^2>3/2$.  Recalling from equation
(\ref{STEADYREQ}) that a fully self-similar inflow/outflow solution would
have $(\ra/\ro)^2=(4-2q)/(3-2q)$, we note that this implies such solutions 
are only possible when $q>1/2$.  From equation (\ref{VPLIMIT}), steady
inflow/outflow solutions would need 
$\vp/(\Omega\ro)<\left[\left(2q-1\right)/(3-2q)\right]^{1/2}$;  since the 
maximum value of the right-hand side is one (for $q=1$)
the asymptotic axial wind speed for such solutions
could not exceed the Kepler speed at the launch point $\Omega\ro$.


\subsection{Self-similar steady wind equations and 
nondimensionalization}\label{SSEQS}

Section \ref{IDEQS} presents the general ideal MHD equations.  Here, we
specialize to the case of self-similar flows in which all the flow 
variables are power-laws in the spherical radius $r$, and in particular 
all velocities
are required to follow the Kepler-law behavior $v\propto r^{-1/2}$.
With this {\it ansatz}, we write the functional forms of the density, 
magnetic flux, field-line rotation rate, magnetic-to-mass flux ratio, 
total specific angular momentum, and Bernoulli parameter as

\begin{equation}\label{DENSITY}
\rho\equiv \rho_1 \left({r\sin\theta\over R_1}\right)^{-q} n(\theta),
\end{equation}

\begin{equation}\label{FLUX}
\Phi\equiv \Phi_1 \left({r\sin\theta\over R_1}\right)^{(3-q)/2} \phi(\theta),
\end{equation}

\begin{equation}\label{ROTATION}
\Omega\equiv \Omega_1 \left({r\sin\theta\over R_1}\right)^{-3/2} 
\omega(\theta),
\end{equation}

\begin{equation}\label{BETA}
\beta\equiv \beta_1 \left({r\sin\theta\over R_1}\right)^{q/2} b(\theta),
\end{equation}

\begin{equation}\label{ANGMOM}
J\equiv J_1 \left({r\sin\theta\over R_1}\right)^{1/2} \ell(\theta),
\end{equation}

and

\begin{equation}\label{BERNPAR}
\cale\equiv \cale_1 \left({r\sin\theta\over R_1}\right)^{-1} \epsilon(\theta).
\end{equation}

In the above expressions, we have anticipated the asymptotic
($\theta\rightarrow 0$) functional form of the flow variables by
explicitly including the appropriate power-law dependence in
$\sin\theta$ in each definition.  So that each quantity is normalized
by its respective asymptotic value on $R=r\sin\theta=R_1$ (for
$\theta\rightarrow 0$, $r\rightarrow \infty$, ), we set $n(0)=1$,
$\phi(0)=1$, $\omega(0)=1$, $b(0)=1$, $\ell(0)=1$, and $\epsilon(0)=1$.
\footnote{For a more general radially 
self-similar flow -- not requiring cylindrical
asymptotics -- we could use the same form of the equations and instead 
normalize by fixing the values of $n$, $\phi$, $\omega$, etc. on an
arbitrary $\theta=\theta_1$, $r=r_1$.}  
As introduced in  \S\ref{CYLEQS}, we define the constants 
\begin{equation}
m\equiv \ma^{-2}\big|_{\theta\rightarrow 0}={\beta_1^2\rho_1\over 4 \pi} ,
\end{equation}
\begin{equation}
j\equiv {J\over \Omega R^2}\bigg|_{\theta\rightarrow 0}= {J_1\over \Omega_1 R_1^2},
\end{equation}
and
\begin{equation}
e\equiv {-\cale \over \Omega^2 R^2}\bigg|_{\theta\rightarrow 0}
={-\cale_1\over \Omega_1^2 R_1^2};
\end{equation}
for cold flow originating at $R_0$ in a Kepler-rotating disk,
$j=(\ra/R_1)^2$, $e=(3/2) (R_0/R_1)^2$, and $m$ is given by equation
(\ref{MACYL}). It is also convenient to define the constant 
\begin{equation}\label{HDEF}
h\equiv {\vpone^2\over (\Omega_1\rl)^2} =1-2e - {(j-1)^2\over(1-m)^2}
\end{equation}
where $\vpone$ is the poloidal speed on the streamline asymptotic to
$\rl$, and the final equality is derived from the asymptotic cylindrical
solution (see eq. \ref{VPCYL}).  

From \S\ref{IDEQS}, the functions 
$\Omega$, $\beta$, $J$, and $\cale$ are all field
line invariants, such that $\Omega=\Omega_1$ on $\Phi=\Phi_1$, etc.  From
expressions (\ref{FLUX}) - (\ref{BERNPAR}), this implies that radially
self-similar flows have
\begin{equation}\label{STREAMINV}
\omega=\phi^{-3/(3-q)},\ b=\phi^{q/(3-q)},\ 
\ell=\phi^{1/(3-q)},\ {\rm and}\  \epsilon=\phi^{-2/(3-q)}
\end{equation}
for all $\theta$.

To derive an equation for the remaining flow function $n(\theta)$,
we use the self-similar expressions (\ref{FLUX}) - (\ref{BERNPAR}) in the
equations (\ref{BPDEF}), (\ref{BETADEF}), and (\ref{UPEQ}) to obtain
\begin{equation}\label{SSBERN}
{h Q^2(\theta)\over (b n)^2} 
= 2\left({\ro\over\rl} \right)^3 \sin\theta - 
2 e \epsilon + \omega^2
-\left({j\ell -\omega\over 1 - m n b^2  } 
\right)^2.  
\end{equation}
Here,
\begin{equation}\label{QDEF}
Q^2(\theta)\equiv \phi^2\sin^2\theta + 
\left(\phi\cos\theta + {2\over 3-q} \phi'\sin\theta
\right)^2,
\end{equation}
where we have set the field line rotation speed $\Omega_0 \ro$ 
at the footpoint $\ro$ equal to the Kepler speed $\sqrt{G M/\ro}$,
and we drop the thermal energy term.  Equation (\ref{SSBERN}) is a
quartic equation for $n(\theta)$ in terms of $\phi(\theta)$, 
$\phi'(\theta)\equiv d\phi/d\theta$, $\theta$, and constant parameters.

To complete the set of wind equations, we must restate the 
Grad-Shafranov equation (\ref{GSEQ}) in terms of the reduced 
self-similar functions defined in equations (\ref{FLUX}) - (\ref{BERNPAR}).
The result is a (somewhat complicated) second-order O.~D.~E. which is 
linear in $\phi''(\theta)$;  the full expression is given in \S\ref{SSGSAPP} 
of the Appendix.

As described in \S\ref{WINDASY}, a self-consistent $r$-similar wind
solution must either become effectively hyperbolic near the pole (with 
parameters
tuned such that the corresponding critical transition through the modified
fast MHD point is made
smoothly), or else if it remains effectively elliptic, the solution must match
some physically realistic boundary conditions at the pole (again, by
appropriately tuning the input parameters).  In the present work, we
satisfy this requirement by matching explicitly to an analytic
solution near the pole.  We choose the $R$-similar cylindrical
solution introduced in \S\ref{CYLEQS} as the $\theta=0$ boundary
condition for our $r$-self-similar solutions.  As described earlier,
this limits our choice of the scaling parameter to $q<1$.
Additional requirements on the asymptotic solution arising from this
choice of polar boundary condition are described in \S\ref{ASYAPP}.

For the cylindrical asymptotic solutions of this work, $\ma$ and $\mf$
become constant, and $|v_\theta|/v_p=\sin\theta$, which approaches zero
near the pole -- hence $|v_\theta|/v_{\rm A}\rightarrow 0 $ near the
pole, and solutions remain effectively elliptic.  Furthermore, we have
found that $\mf$ remains $<1$ throughout the flow for all the
cylindrically-collimated solutions we obtain, so these models would be
elliptic when considered as solutions of the full two-dimensional
equations, as well.  Thus, the solutions obtained in this work must
negotiate only the first critical point, requiring a smooth transition
through $\ma=1$.  The conditions that must be met at the Alfv\'en
point are detailed in \S\ref{ALFAPP}.

The numerical solutions in this work are obtained by initiating the
integrations at the pole, and tuning the input parameters until a
smooth \alf\ transition is obtained.  For some solutions, we find it
convenient to match onto the sub-Alfv\'enic part of the flow with a
numerical integration that starts at the equator; the two solutions
then overlap in the sub-Alfv\'enic region.  Once the input parameters
have been found such that the flow integration initiated at the pole
negotiates the \alf\ critical
transition, such matching solutions require no additional choice of
input parameters at the equator 
to be fully-specified.  When using the shooting
method and starting from the equator, however, it is necessary to
search for the correct initial value of $\phi'$ at the equator such
that a good match is obtained.  The governing equations become singular
at the equator for a cold flow (since $n\rightarrow\infty$ as
$v\rightarrow 0$); the limiting behavior of the equations and
implications for the equatorial boundary condition are discussed in
\S\ref{EQAAPP}.


\section{Magnetocentrifugal disk wind solutions}\label{SOLS}

In this section, we present solutions we have obtained for $r$-similar cold
MHD flows that meet cylindrical ($R$-similar) boundary conditions at the
pole.  As explained in \S\S\ref{CYLEQS}-\ref{SSEQS}, the solutions we obtain
are described by $q$ (the scaling parameter for which $\rho\propto r^{-q}$
and $B\propto r^{-(1+q)/2}$),
 and the two parameters $j=(\ra/\rl)^2$ and $e=(3/2)(\ro/\rl)^2$,
where $\ro$ is the radius of a streamline footpoint in the disk, $\ra$ is
the cylindrical \alf\ radius, and $\rl$ is the asymptotic cylindrical 
radius.  For our numerical solutions, we fix $q$ and $\ro/\rl$ and then vary 
$\ra/\ro$
until the conditions for a smooth \alf\ transition are satisfied (see
\S\ref{ALFAPP}).  We thus obtain a two-parameter family of solutions.  

\begin{figure}
\plotone{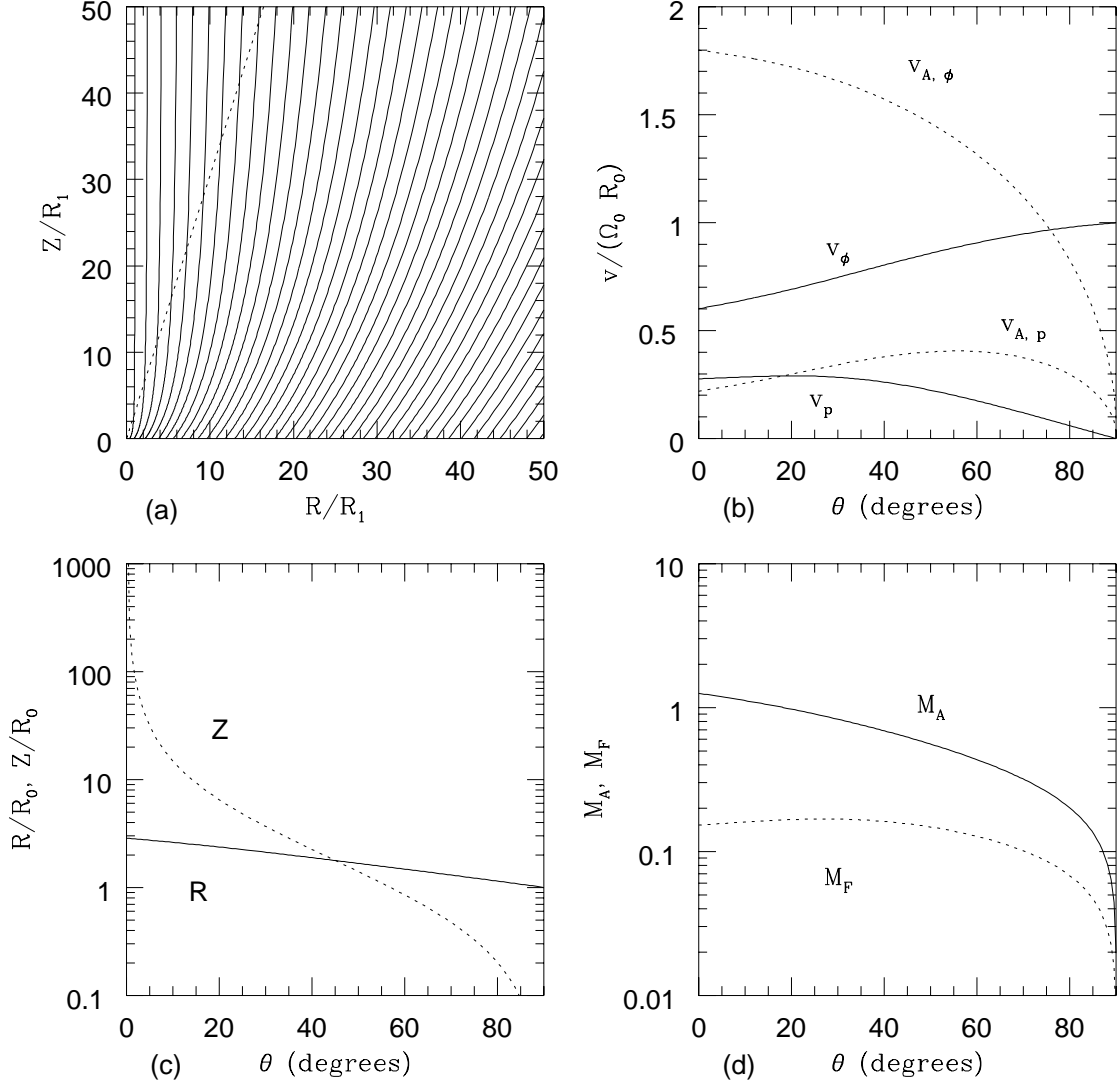}
\caption{
Flow solution for q=0.75, $\ro/\rl=0.35$, $\ra/\ro=2.41$.
(a) Streamlines (solid curves) for equal increments in $\dot M$.  The
innermost streamline originates at $R/\rl\equiv\ro/\rl=0.35$ on the
abscissa and asymptotes to $R/\rl=1$ at infinite $Z$;  the corresponding 
flow makes an \alf\ transition at $R/\rl\equiv\ra/\rl=2.41$. The dashed line
shows the locus of the \alf\ surface.  (b) Fluid speeds (solid curves) and
\alf\ speeds (dashed curves) of poloidal and toroidal flow/field components,
in units of the Kepler speed $\Omega_0\ro$ at the footpoint of a streamline,
as a function of angle $\theta$ with respect to the pole.  (c)  Streamline
radial distance (solid curve) and height above the disk (dashed curve) 
in units of the footpoint radius $\ro$ at the equator.  (d)  \alf-mode
 (solid curve) and fast-mode (dashed curve) Mach numbers for the flow,
$\ma=\vp/\vap$ and $\mf=\vp/\va$. 
}
\end{figure}

An example of a solution for $q=0.75$ and $\ro/\rl=0.35$ 
is portrayed in Figure 2a-d.  We find that the solution requires 
$\ra/\ro=2.412438$ in order to make a smooth \alf\ transition;  
the \alf\ surface lies at $\theta=18^\circ$ with respect to the pole.  
From Figures 2a and 2c, collimation becomes very rapid after the \alf\ 
transition has been made.  
From Figure 2c, it is apparent that there is no radial ($R$) oscillation of the
streamlines;  instead, in this and our other solutions the cylindrical 
radius $R$ of any streamline secularly increases with height above the 
disk.  From Figure 2b, notice that
near the disk (at large $\theta$) $\vaphi$ rises above $\vap$ even before 
the \alf\ transition is made.  This effect occurs in all of the solutions
we have obtained.  Thus, the acceleration in these flows may be thought of as
largely due to the gradient in the toroidal field pressure,
rather than primarily due to the centrifugal force in
a rotating, nearly rigid, poloidal magnetosphere (cf. e.g. Spruit 1996).
From Figure 2d, notice that the flow is sub-fast-MHD ($\mf<1$) 
throughout.  This is also true for the rest of the solutions we
have found.  Another feature of Figure 2b which holds for the other flow 
solutions as well is that the asymptotic rotation speed exceeds the 
asymptotic poloidal (axial) speed.  The slight decrease in poloidal speed 
$\vp$ near the pole is also a general feature of our solutions.  

\begin{figure}
\plotone{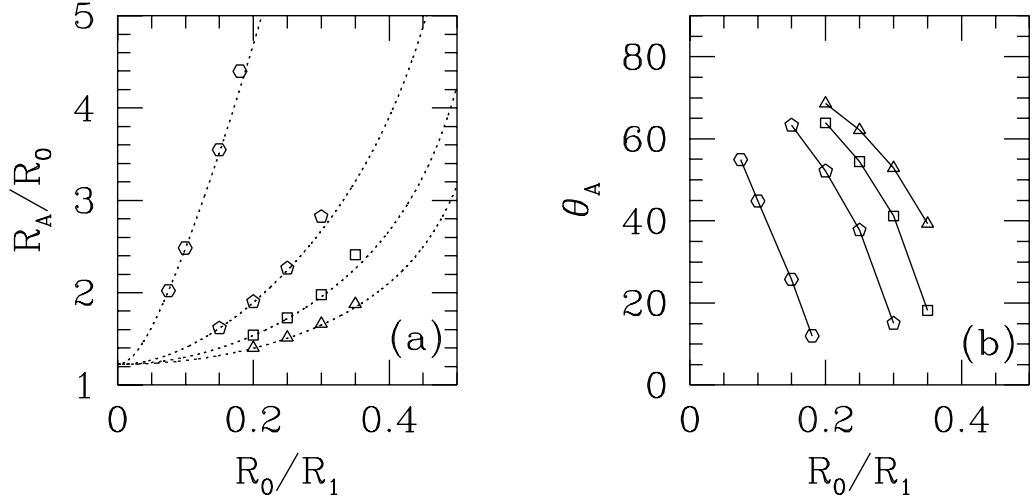}
\caption{
(a) Relation among parameters $\ra/\ro$, 
$\ro/\rl$, and scaling power $q$, for several outflow solutions.
$\ro$, $\ra$, and $\rl$ are the respective cylindrical radii of the
footpoint, \alf\ transition, and asymptotic lateral expansion of each 
streamline; 
$q=-{\partial \log\rho/\partial\log r}=-(1+2{\partial\log B/\partial \log r})$.
Triangles show solutions with $q=0.5$; squares $q=0.75$; pentagons
$q=0.9$; hexagons $q=0.99$.  The dashed curve shows the lower boundary for
cylindrical asymptotic solutions to exist, i.~e. the locus where 
$\vp=0$ for a self-similar cylindrical flow of given $q$ (see eq. \ref{JMIN}).
(b) Angle of the locus of the \alf\ surface with respect to the pole, for the 
solutions in (a) (solid lines connect the solution points of given $q$).
}
\end{figure}

While the general properties of the flow depicted in Fig. 2 is
representative of all the solutions we have found, the quantitative
characteristics of course differ for each solution.  Figure 3a shows
the relationship among the three parameters $q$, $\ro/\rl$, and
$\ra/\ro$ that characterise flows which become asymptotically
cylindrical, for a number of solutions we have obtained which
successfully make a smooth \alf\ transition.  Numerically, $\ra/\ro$ is
obtained as an eigenvalue for any ($q$, $\ro/\rl$) pair. From Figure
3a, notice that the solutions all hug near the limiting locus ($\vp=0$) for
cylindrical asymptotic solutions to exist.
For fixed similarity scaling power $q$, solutions with relatively more
angular momentum (large $\ra/\ro$) collimate relatively close to their
footpoints (large $\ro/\rl$), and also have their \alf\ radii
relatively closer to the asymptotic cylindrical radius (large
$\ra/\rl$).  For fixed ratio $\ra/\ro$ of the \alf\ cylindrical radius
to the streamline footpoint radius, an increase in the central
concentration of the magnetic flux and density (i.e. an increase in
$q$) implies a smaller value of $\ro/\rl$ -- i.e. the solution expands
to a larger cylindrical radius before collimating.

In Figure 3b, we show the corresponding value of $\theta_\rma$, the angular
distance of the \alf\ surface from the pole, for the
solutions shown in Figure 3a. For fixed $q$, solutions which collimate
relatively close to their footpoints ($\ro/\rl$ large) make their
\alf\ transitions relatively near the pole ($\theta_\rma$ small).  
For fixed $\ro/\rl$, the streamlines for different values of $q$ follow
very nearby paths, and the variation of the fluid speeds with angle $\theta$
is also quite similar for the solutions with differing $q$.  On the other
hand, the variation of the \alf\ speeds with $\theta$ differs significantly
for solutions with different $q$ and the same $\ro/\rl$;  as $q$ increases,
the \alf\ speed at a given $\theta$ increases, thereby shifting the \alf\ 
point closer to the pole. 

\begin{figure}
\plotone{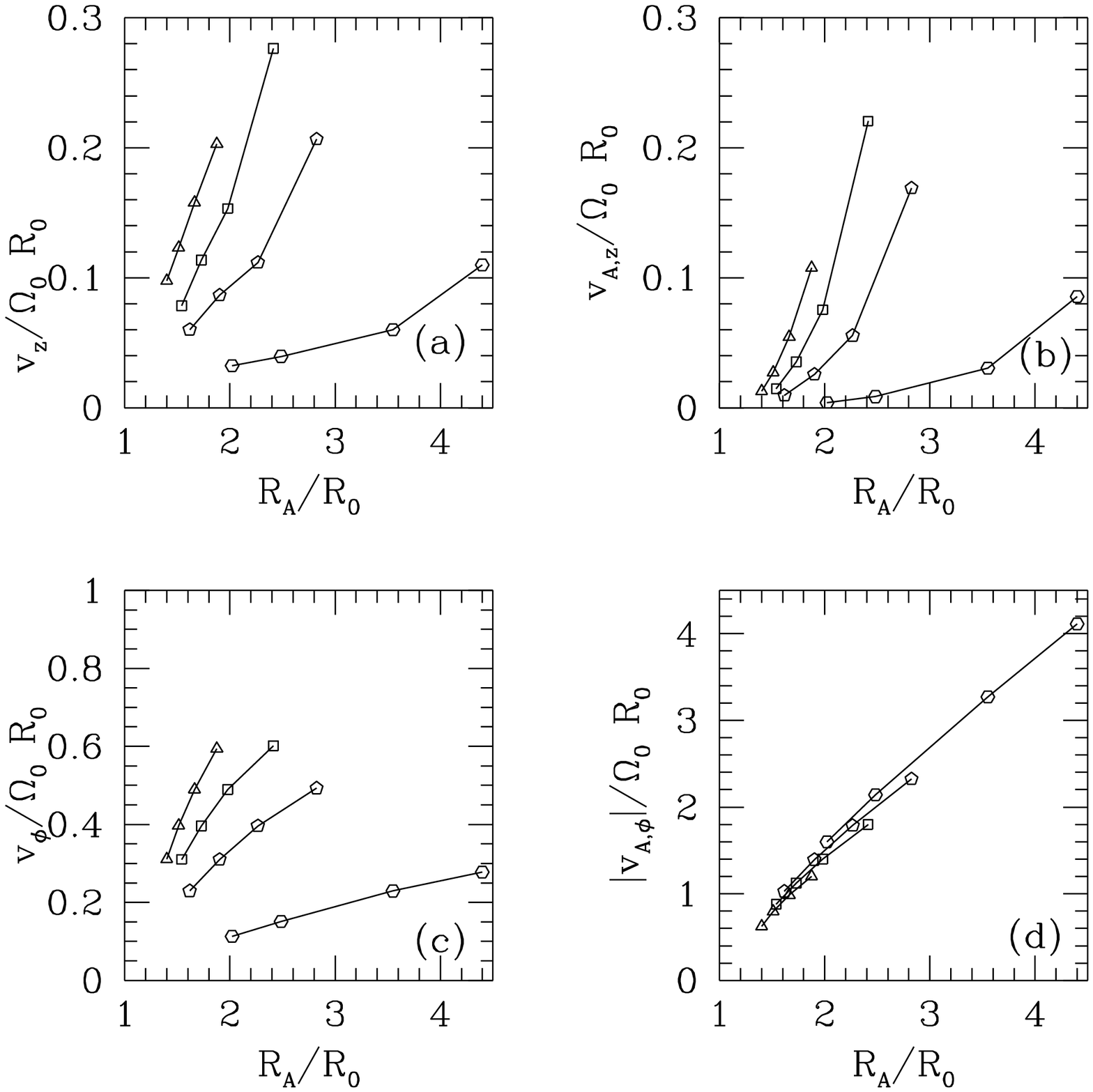}
\caption{
Asymptotic fluid and \alf\ speeds for full $r$-similar
solutions with various parameters, in units of the Kepler speed at the
streamline footpoint.  Triangles show solutions with $q=0.5$; squares
$q=0.75$; pentagons $q=0.9$; hexagons $q=0.99$.  Solid lines connect
the solution points.  (a) Poloidal (axial) speed. (b) Poloidal \alf\
speed. (c) Toroidal speed. (d) Toroidal \alf\ speed.
} 
\end{figure}

The asymptotic flow and \alf\ speeds vary with the flow parameters $q$, 
$\ra/\ro$, and $\ro/\rl$.  In Figure 4, we plot the values of
these asymptotic speeds for the same set of solutions as in Figure 3.
From Figure 4a, notice that for a given similarity scaling $q$, the 
asymptotic axial fluid speed increases as the \alf\ point moves further
from the streamline footpoint ($\ra/\ro$ increases).  At fixed $\ra/\ro$,
solutions with increasingly concentrated magnetic flux (increasing $q$)
have lower asymptotic $v_z$.  From Figures 4b and 4c, the same trends hold
for the poloidal \alf\ speed and the toroidal flow speed.  Figure 4d shows
that the asymptotic toroidal \alf\ speed increases with $\ra/\ro$ for
fixed $q$.  At fixed $\ra/\ro$, the toroidal \alf\ speed increases very 
slightly with the degree of flux concentration (increasing $q$). 
Note that none of the solutions displayed here have $\ra/\ro$ sufficiently
small to correspond to the case of a steady, self-similar inflow driven
by the outflow at given $q$ (cf. eq. \ref{STEADYREQ}).  While we have 
identified some such solutions, the extremely low asymptotic poloidal speeds
that they yield makes them unlikely candidates to model real outflows.

\begin{figure}
\plotone{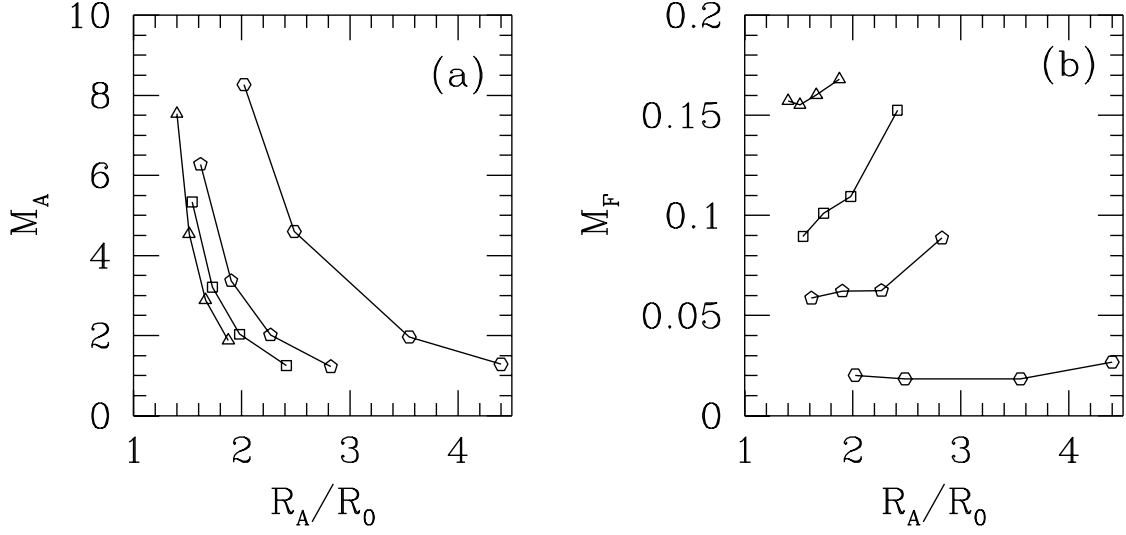}
\caption{
Asymptotic \alf\ and fast MHD
Mach numbers $\ma\equiv \vp/\vap$ (a) and $\mf\equiv \vp/\va$ (b) for
$r$-similar solutions with various parameters.  Triangles show
solutions with $q=0.5$; squares $q=0.75$; pentagons $q=0.9$; hexagons
$q=0.99$.  Solid lines connect the solution points.  
}
\end{figure}

Figure 5a,b shows how the \alf\ and fast-MHD Mach numbers
$\ma$ and $\mf$ for the asymptotic limit of the
$r$-similar flow varies with the solution parameters.  For fixed
similarity scaling $q$, $\ma$ decreases with increasing $\ra/\ro$,
while $\mf$ slightly increases with $\ra/\ro$.  At fixed $\ra/\ro$, an
increase in the concentration of the flow (larger $q$) implies a
larger asymptotic $\ma$, but a smaller asymptotic $\mf$. On the other
hand, for fixed ratio of initial to final radius of a streamline
$\ro/\rl$ (not shown), an increase in $q$ implies both smaller $\mf$
and $\ma$, asymptotically.

The flow solutions we have obtained are in general magnetically dominated
in the asymptotic regime, in the sense that for most of the solutions,
the energy, momentum, and angular momentum fluxes of carried by Maxwell
stresses exceed the respective kinetic fluxes.  
These flux ratios are given, respectively, by
\begin{equation}
{E_K\over E_M}= {|{\bf v}|^2 \ma\over 2\Omega R |\vaphi|},
\end{equation} 
\begin{equation}\label{MOMFLUX}
{P_K\over P_M}= ({1\over 2} \mf^{-2} -\ma^{-2} )^{-1},
\end{equation} 
and
\begin{equation}
{J_K\over J_M}= {\vphi\ma\over |\vaphi|}.
\end{equation} 
(Note that the small electric contribution $(1/2) \ma^{-2} (\Omega R/c)^2$ to
the denominator of equation (\ref{MOMFLUX}) has been dropped.)
Figure 6 shows the asymptotic values
of these flux ratios for the same solutions as discussed above.  A few
cases with small $q$ and $\ra/\ro$ have kinetic angular momentum flux 
exceeding the magnetic angular momentum flux, but otherwise the solutions
are magnetically dominated.  

\begin{figure}
\plotone{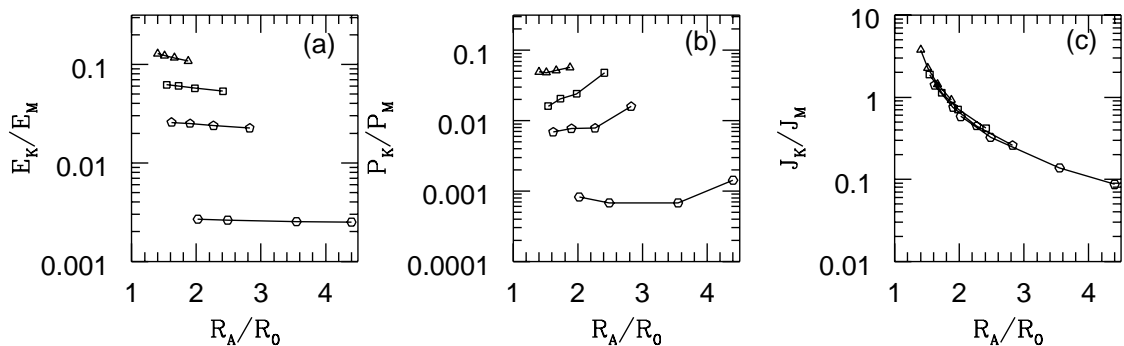}
\caption{
Asymptotic ratios of kinetic to magnetic 
energy flux (a), momentum flux (b), and angular momentum flux (c) for
$r$-similar solutions with various parameters.  Triangles show
solutions with $q=0.5$; squares $q=0.75$; pentagons $q=0.9$; hexagons
$q=0.99$.  Solid lines connect the solution points.  See text for definitions
of fluxes.
}
\end{figure}

\vfil\eject

\section{Summary and Discussion}

The calculations presented in this paper explore the proposal that
observed narrow optical jets from PMS stars faithfully represent an inherent
strong collimation in both density and velocity of the winds which are
magnetocentrifugally driven from these young star/disk systems.  To this
end, we have developed models of MHD disk winds in which all streamlines 
asymptote to lie on the surfaces of a series of nested cylinders, at large
distances from the disk.  To make these calculations tractable, we work 
within the framework of radially self-similar flows, for which the density 
$\rho$, all magnetic field strengths $B$, and all velocities $v$ have 
power-law scaling in the spherical radius $r$: $\rho\propto r^{-q}$, 
$B\propto r^{-(1+q)/2}$, and $v\propto r^{-1/2}$, for arbitrary $q$.  
The shape of wind 
streamlines and the dependence of all flow quantities on $\theta$
are then calculated self-consistently from the steady-state MHD equations.
The $r$-self-similar ansatz is a convenient mathematical 
idealization, and also may serve as a good characterization of 
the possible flows well away from the boundaries of a disk with large 
dynamic range.  Along a ray of $\theta=const.$, the decrease of the dynamical
variables ($\bf v$, $\bf B$, $\rho$) with $r$ occurs because the collimation 
of the flow brings streamlines from successively more distant regions of the 
disk to cross the ray.  Since the farther regions of the disk are 
characterized by lower velocity, and by assumption lower density $\rho$ and 
magnetic field strength $B$, the wind will reflect these decreases 
correspondingly.  Note in particular that the present mathematical
formalism allows for fast flows at large distance $r$ from the source even
though $v\propto r^{-1/2}$:  if $v_p$ increases rapidly with decreasing 
$\theta$ (toward the $\hat z$-axis), then a ``jetlike'' flow can exist
along the pole.

We begin our analysis, in \S 2.2, 
by presenting an exact, analytic family of solutions for
rotating, cylindrically-symmetric, axial ($v_R,B_R=0$) MHD flows which are 
self-confined
by gradients in the toroidal magnetic field.  All flow variables are 
independent of the distance along the flow axis $z$, and have power-law 
dependence $\rho\propto R^{-q}$, $B_\varphi\propto B_z\propto R^{-(1+q)/2}$, 
$v_\varphi\propto v_z\propto R^{-1/2}$
on the cylindrical radius $R=r\sin\theta$.  The 
three-parameter family of solutions can be described by $q$ together with 
$j$ and $e$, where the latter two respectively are 
the specific angular momentum, and the Bernoulli constant 
(equivalent to the fluid energy in a rotating frame), in units of 
local rotation rate of a magnetic field line $\Omega$ and cylindrical 
radius $R$.  These parameters are natural to use when connecting
to a flow from a disk in Keplerian rotation.  This fully-analytic family of
shearing, 
rotating, axial winds includes both sub- and super- fast MHD flows, i.e. 
$v_z<$ or $>\va$.  
The presence of a toroidal field $\vaphi\ne 0$ permits confinement of
rotating flows in which the magnetic pressure decreases to larger $R$ -- 
i.e., flows in which the outward centrifugal force and outward pressure
gradient forces are balanced by an inward magnetic tension (``hoop stress'').
In order for this to work, however, the gradient in the magnetic field cannot 
be too steep: for self-similar flows, cylindrical collimation is only
possible if $B$ drops with $R$ no faster than $R^{-1}$, i.e. $q<1$.

We next, in \S 3, take the family of rotating, shearing
cylindrical flows as the set of
desired asymptotic boundary conditions far from the disk, and ask what
sort of magnetocentrifugally-driven disk winds can achieve this fully
collimated state.  We present a range of solutions for various $q$,
$e=(3/2)(\ro/\rl)^2$ and $j=(\ra/\rl)^2$, where $\ro$ is the position of a
streamline's footpoint in the disk, $\ra$ is the \alf\ radius, and $\rl$ is the
asymptotic radius.  Note that the self-similarity of the flow makes $e$ and $j$
the same for all streamlines.  For each $q$ and $e$ there is a unique $j$ for
which the flow passes smoothly through the \alf\ point; thus the
family of $r$-similar disk winds which become cylindrically collimated
is labeled by two parameters.  The asymptotic characteristics shared
by all the self-collimating wind solutions presented 
are (1) $v_z/\Omega \ro\sim$
few $\times 0.1$, (2) $v_\varphi/\Omega \ro\sim$
few - several $\times 0.1$,  (3) $\vaphi/\vap>>1$, and (4) $v_z/\va<1$.
Here, $\Omega$ is the Kepler speed in the disk at the streamline's footpoint
$\ro$.
In particular, note that unlike in most previous MHD disk wind solutions
(which, however, have unconstrained asymptotic states), 
these flows do not accelerate to final axial speeds comparable to or
larger than the escape
speed from the potential well where they originated.  
Thus, for example, if the innermost streamline originates near the surface
of a PMS solar-type star, then based on the present solutions 
the maximum outflow speed along the central
axis of the ``jet'' would only be a few to several tens of $\kms$, much lower 
than the optical Herbig-Haro jets observed to reach a few hundred $\kms$ 
speeds.
Intuitively,
the result (4) may be understood as follows:  a cold MHD flow must leave the 
disk
at an angle at least $30^\circ$ from the vertical.  In order to be fully
redirected upwards, $\vaphi$ must be $\gtrsim \vp$ to accomplish 
the refocusing.
But after the flow has been redirected, it is left in sub-fast-MHD state.

Because the fully-collimated disk wind solutions we have found have relatively
low final speeds, we believe that observed optical jets from YSOs are 
unlikely to be manifestations of these or similar flows.  Among the many
possible alternative models, we list a few:  

\par\noindent{\bf(1)}  The wind responsible for the observed 
jet originates in the disk, and its final velocity 
collimation is large but less extreme than the present solutions.  
\cite{con94} have identified solutions with $q$ slightly greater than one
which continue expanding laterally (in $R$) up to the limit of their 
integration in $z$, and have $\vp/\Omega\ro>1$.  It would be interesting to
compare the density and momentum distributions in such solutions with 
observed jets.

\par\noindent{\bf(2)} A well-collimated MHD disk wind like those computed in 
this paper comprises an unseen neutral wind which surrounds the
optical jet and enforces its observed collimation; the wind would also
help drive outflows by directly sweeping up ambient material far from
the jet.  The jet itself must originate inside of the collimating
wind, and have a more efficient acceleration mechanism (perhaps
involving relatively strong poloidal magnetic fields to achieve the
magnetic ``propeller'' effect).  Since in this case the jet is not
required to self-collimate its velocity vectors, the flow which produces the 
jet could emerge from near the star in a relatively isotropic fashion.

\par\noindent{\bf (3)} As in the model of Shu et al (1994a,b), 
\cite{naj94}, and 
\cite{shu95}, the whole of the wind originates in a narrow region of the disk 
near the star and flows outward with nearly radial poloidal streamlines
and speeds comparable to the stellar escape speed for the whole flow.  
At large distances from the star,  the velocity field is still nearly
radial with only logarithmic streamline 
collimation due to toroidal field stresses.
In this case, however, the density appears much more cylindrically collimated
than the velocity field, and the optical jet may represent just the densest 
part of the wind near the axis.

Since the models presented herein are not ``global'' in the sense of physically
matching to interior and especially exterior ambient conditions, it is more
difficult to assess the possible relationship between such winds and observed
molecular outflows.  A real disk has a real inner and outer edge, and thus 
there must be a first and last wind streamline.  If we arbitrarily truncate
the source of the self-similar wind at the inside and outside disk edges
$\ro=R_*$ and $\ro=\rd$,
then the first and last streamlines would exit the
disk at a finite angle\footnote{Note that to maintain force balance near
the outer 
disk edge there would in fact need to be additional wind streamlines filling
the region toward the equatorial plane.}
 and asymptote to helically wrap around the surfaces of
cylinders of inner and outer radii $R_{\rm C}=(\rl/\ro)R_*$ and 
$\rw=(\rl/\ro)\rd$, respectively.  If we assume a central star of mass 
$M_\star$ 
and total mass-loss rate $\dot M_{\rm W}$ in the wind between $R_*$ and $\rd$, then
we can introduce dimensional scales for the density, magnetic field, and 
velocity to write the asymptotic profile in the wind at cylindrical 
distance $R$ from the axis as
\begin{equation}
\rho=\rho_{\rm W} \left({R\over \rw}\right)^{-q},
\end{equation}
\begin{equation}
{\bf B}={\bf B_{\rm W}} \left({R\over \rw}\right)^{-(1+q)/2},
\end{equation}
and 
\begin{equation}
{\bf v}={\bf v_{\rm W}} \left({R\over \rw}\right)^{-(1/2)}.
\end{equation}
Here,
\begin{equation}
\rho_{\rm W}= 4.5\times 10^5 \cm^{-3} \mu_H 
\left({\dot M_{\rm W}\over 10^{-7}\msun\yr^{-1} }\right)
\left({M_\star\over\msun }\right)^{-1/2}
\left({\rd\over 100\au }\right)^{-3/2}
\left({\ro\over\rl }\right)^2
\left({v_z\over \Omega \ro}\right)^{-1}
N_q
\end{equation}
is the density on the outermost wind streamline, 
\begin{eqnarray}
{\bf B_{\rm W}}&&=\left({v_{{\rm A}, z}\over\Omega\ro},
{\vaphi\over\Omega\ro}\right)
\times
\nonumber\\
&&
0.92{\rm mG} 
\left({\dot M_{\rm W}\over 10^{-7}\msun\yr^{-1} }\right)^{1/2}
\left({M_\star\over\msun }\right)^{1/4}
\left({\rd\over 100\au }\right)^{-5/4}
\left({\ro\over\rl }\right)
\left({v_z\over \Omega \ro}\right)^{-1/2}
N_q^{1/2}
\end{eqnarray}
gives the magnetic field components on the outermost streamline,
and
\begin{equation}
{\bf v_{\rm W}}=\left({v_z\over\Omega\ro},{v_\varphi\over\Omega\ro}\right)
\times 3.0\kms
\left({M_\star\over\msun }\right)^{1/2}
\left({\rd\over 100\au }\right)^{-1/2}
\end{equation}
gives the components of the wind velocity on the outermost streamline;
\begin{equation}
N_q\equiv{3/2-q\over 1-(R_*/\rd)^{3/2-q}}
\end{equation}
is a normalization constant.
The values of $\ro/\rl$, and asymptotic 
$v_z/\Omega\ro$, $v_\varphi/\Omega\ro$,
$v_{\rm A,z}/\Omega\ro$, and $\vaphi/\Omega\ro$ are shown for various models
in figures 3d and 4a-d of \S3.

The pressure in the wind is dominated by $\bphi^2/8\pi$.  While the very 
innermost part of the parent cloud 
core may have comparable pressure (at density 
$\sim 10^8\cm^{-3}$), the outer cloud core ambient medium 
will generally have pressure far
below that of the wind.   Thus, either the wind as
a whole or its surface layers 
would have to expand laterally until the pressure matches the 
ambient medium (which event occurs depends in part on the stability of the 
wind, a question under current investigation).  

If the wind expands as a whole,
keeping the mass load and poloidal speed on streamlines unchanged, then
in the limit of negligible ambient pressure the wind would fill all $4\pi$
steradians. The far-asymptotic solution can be computed similarly to the
calculation outlined in \cite{shu95}, if we assume that the flow
streamlines adjust their latitudes $\theta$ at each $r$ so that the force
associated with the hoop stress (i.e. the gradient in $(R\bphi)^2$) is zero.
The resulting streamlines and density contours have qualitatively similar
behavior to the \cite{shu95} results for a wind originating from a narrow
region in the disk near the star.  In particular, the distribution of wind
momentum flux with angle has $d\dot P/d\cos\theta\sim C(r)\ro^{(1-q)/2}
/(\sin\theta)^2$ for $C(r)$ a slowly-varying function, and $\ro$ the footpoint
radius of the streamline that passes through $\theta$ at that $r$.  For $q$
near unity, this is nearly the same momentum distribution as that used
by \cite{li96c} in their calculations of the lobe shapes and line profiles
for swept-up molecular shells, so their results would carry over to the case 
of ``fully expanded'' disk winds. As \cite{li96c} showed, a momentum
distribution near $d\dot P/d\cos\theta\propto 1/(\sin\theta)^2$  
and toroidal mass distribution $\rho_{\rm amb}\propto \sin^2\theta$ in the
ambient medium well reproduces the parabolic outflow shell 
shapes and line profile wings $dM_{sh}/dv_{sh} \propto v_{\rm sh}^{-1.8}$ 
characteristic of many observed outflows.
The coincidence in momentum distribution between the present case and the 
X-wind models occurs because here, the 
lower poloidal speeds on the outer disk streamlines are offset by
a relatively greater mass load, compared to the X-wind models.
Because winds from the outer portions of disks would have lower speeds but
higher densities than the corresponding X-wind with the same momentum flux,
the two cases could potentially be distinguished by the emission properties
of the region where the wind impacts the ambient cloud.  For example, finding
H$_2$ emission at low latitudes in outflow shells would argue against 
disk-wind models for sweeping up the shell, because a shocked low-speed 
wind would not reach the few-thousand Kelvin temperature required.

If instead of the wind expanding as a whole, we assume that only the surface 
layers expand to match the pressure in the ambient medium, then
much of the wind could retain its axial poloidal velocity field.  
The distribution of axial
momentum flux in this interior portion would obey 
$d\dot P/dR\propto R^{-q}$.  If this inner wind sweeps up
a shell from a surrounding cloud core with initial density distribution 
$\rho_{\rm amb}\propto z^{-a}$, then the distribution of mass with velocity
$v_s$ in the shell would obey 
$dM_{sh}/dv_{sh}\propto v_{sh}^{-4/(1+q)+a(1-q)/(1+q)}$.  
For $q$ near unity, or
$a$ near $2$ (as for the \cite{shu77} singular isothermal sphere), the power
in the distribution is
 near $-2$, in good agreement with the $-1.8$ dependence cited by 
\cite{mass92} for the high-velocity part of observed outflow line profiles.  
Without a model for how the wind on the outer streamlines expands and 
interacts with the ambient gas, however, it is impossible to tell whether 
such a ``partially expanded'' collimated disk wind would be consistent with 
observed outflow shapes and the distribution of low-velocity gas.

In summary, we believe that highly-collimated fast flows as defined by 
such famous jets as HH30, HH34, or HH111 are unlikely to represent the
whole story of the primary winds from PMS star/disk systems.  The models
of this paper, though by no means an exhaustive survey of jet 
production possibilities,  
show that it may be difficult to generate strong {\it velocity}
collimation and fast outflow self-consistently in a
magnetocentrifugally-launched wind from an extended region of an 
accretion disk.  
The narrow appearance of fast optical jets from young stars
therefore argues in favor of a surrounding wind, either 
from the whole of the disk or from near the star, which sustains the observed
jet collimation and helps to drive molecular outflows on large scales.

\acknowledgements

This work was supported in part by a postdoctoral fellowship from the 
Harvard-Smithsonian Center for Astrophysics.  It is a pleasure to acknowledge
numerous discussions with Ramesh Narayan, Charles Gammie, and 
Jim Stone, as well as helpful comments on the manuscript from Bruce Draine 
and an anonymous referee.

\appendix

\section{Appendix}

\subsection{The Grad-Shafranov equation for $r$-self-similar 
flows}\label{SSGSAPP}

The Grad-Shafranov (GS) equation describes force balance in the direction
perpendicular to the poloidal field lines.  With the adoption of a
radially self-similar form for the flow (eqs. \ref{FLUX} - \ref{BERNPAR}), 
the GS equation becomes a second-order ODE for the reduced flux function
$\phi(\theta)$.  Denoting derivatives with respect to $\theta$ as primes,
this equation can be written
\begin{equation}\label{SSGSEQ}
\phi ''={\cal R\over \cal L}
\end{equation}
where 
\begin{eqnarray}\label{LEFTCO}
{\cal L}&=& (m n b^2 -1)
\left\{
m n b^2\left[ 
2\left({\ro\over\rl}\right)^3\sin\theta + \omega^2 - 2e\epsilon\right]
-h\left({\phi\sin\theta\over n b}\right)^2
\right\} \nonumber \\
	&=& \left[\left({\vap\over \vp}\right)^2  -1\right]
\left({\ro\over\rl}\right)^3{(\va^2-v_\theta^2) \over GM/R}
\end{eqnarray}
and ${\cal R}= {\cal R}_1 + {\cal R}_2 + {\cal R}_3$.
Here, 
\begin{eqnarray}\label{R1EQ}
\lefteqn{
{\cal R}_1 = {3-q\over 4}{{\cal D}(b n)^2\over\phi h\sin^2\theta}\times}
\nonumber \\
& & \left\{ 
(q+1) m n b^2\left[ 
2\left({\ro\over\rl}\right)^3\sin\theta + \omega^2 - 2e\epsilon\right]
-2\left({\ro\over\rl}\right)^3\sin\theta + 
2 {(j\ell)^2 - m n b^2 \omega^2 \over 1 - m n b^2}  
\right\},
\end{eqnarray}

\begin{equation}\label{R2EQ}
{\cal R}_2={\cal D}(1-mnb^2)\phi'
\left(3{\cos\theta\over\sin\theta} +{q+1\over 3-q} {\phi'\over\phi}\right)
\end{equation}

and 
\begin{eqnarray}\label{R3EQ}
\lefteqn{
{\cal R}_3= (1-m n b^2)\left(\phi' + {3-q\over 2}
{\cos\theta\over\sin\theta}\phi\right)
\times}\nonumber\\
& & 
\left\{{h\over 2 (n b)^2}\right.
\left[ \left({2\over 3-q}\right)^2{\phi'\over\phi}\sin\theta
\left(5 \phi\cos\theta+ {2q\over 3-q}\phi'\sin\theta\right)
\left(\phi' + {3-q\over 2}{\cos\theta\over\sin\theta}\phi\right) 
+ {2\over 3-q} \phi'\phi
\right]\nonumber\\
& & \left.-\left({\ro\over\rl}\right)^3\cos\theta -
{2e\epsilon-3\omega^2\over 3-q}{\phi'\over \phi}
+{(j\ell -\omega)(j\ell + 3\omega)\over(3-q)(1-m n b^2)^2}
{\phi'\over\phi}\right\},
\end{eqnarray}
with 
\begin{eqnarray}\label{DDEF}
{\cal D}&=& 
m n b^2 
\left[2 \left({\ro\over\rl}\right)^3\sin\theta + \omega^2 - 2e\epsilon\right]
- {h Q^2 \over (nb)^2} \nonumber \\
	&=&(m n b^2 -1) 
\left[2\left({\ro\over\rl}\right)^3\sin\theta + \omega^2 - 2e\epsilon\right]
+ {(j\ell -\omega)^2\over (1-m n b^2)^2}\nonumber \\
	&=&
 \left({\ro\over\rl}\right)^3{(\va^2 -\vp^2) \over GM/R}.
\end{eqnarray}
The function $Q$, which is proportional to the poloidal field strength, is
defined in equation (\ref{QDEF}).
The constants $e$, $j$, $m$, and $h$ are defined in \S\ref{SSEQS}.
The first two are the free parameters which must be chosen before initiating
an integration, while $m$ and $h$ may be expressed in terms of $e$ and $j$
using the asymptotic cylindrical solution (see eqs. \ref{MACYL} and 
\ref{HDEF}).

\subsection{Asymptotic boundary conditions}\label{ASYAPP}

The equation (\ref{SSGSEQ}) contains terms that scale as $(\sin\theta)^{-2}$ 
and $(\sin\theta)^{-1}$, and this singular behavior near the pole demands
special treatment.  The method is to approximate 
\begin{equation}\label{PHIEXPAND}
\phi(\theta)=1+\phi'(0)\theta + \phi''(0) \theta^2/2
\end{equation}
 near the pole, and
then to expand the governing equation (\ref{SSGSEQ}) in $\theta$ and 
collect terms at each order.
The only part of equation (\ref{SSGSEQ}) that contains
terms of order $(\sin\theta)^{-2}$ is ${\cal R}_1$, and equation
(\ref{MAQUAD}) shows that the expression in curly braces in equation 
(\ref{R1EQ}) contains no order-unity terms.  This verifies that the 
cylindrical solution presented in \S\ref{CYLEQS} is a valid limiting solution 
of the full self-similar wind equations, to lowest asymptotic order.  Next,
we must collect the order $(\sin\theta)^{-1}$ terms in the expansion of 
equation (\ref{SSGSEQ}).
By requiring that the corresponding coefficient be zero, after some algebra
we obtain a linear
equation which we solve for $\phi'(0)$ in terms of $q$, $e$, and $j$.  
Finally, the expansion of (\ref{SSGSEQ}) to order $1$ is linear in 
$\phi''(0)$;  rather than performing the expansion analytically, however,
we numerically solve for the value of $\phi''(0)$, once $q$, $e$, $j$, and
$\phi'(0)$ are set.  After solving for $\phi''(0)$, we choose a 
value for $\theta_1$ (typically $10^{-4}$ is adequate) at which to initiate 
the integration, then evaluate $\phi$ at $\theta_1$ 
via equation (\ref{PHIEXPAND}),
set $\phi'(\theta_1)=\phi'(0) + \phi''(0)\theta_1$, and begin the numerical
integration.

\subsection{Alfv\'en transition constraints}\label{ALFAPP}

As discussed by BP and other authors, equation
(\ref{SSGSEQ}) can become singular if ${\cal L}=0$, which from equation
(\ref{LEFTCO}) occurs if
$\vap=\vp$ (i.~e. $\ma=1$) or $\va=v_\theta$.  For waves propagating along
the $\hat\theta$ direction (i.e. along the spatial direction corresponding to 
the only degree of freedom for a radially self-similar flow), the first case 
corresponds to
an Alfv\'en transition in $v_\theta$, and the second case to a 
fast MHD transition in  $v_\theta$.
To avoid singularities, physical solutions must have ${\cal R}=0$ wherever
${\cal L}=0$;  such points become critical points of the flow, and the
requirement that the flow passes smoothly through these critical points 
constrains the possible values of the parameters entering the solution.
It turns out that for the solutions presented in this work, 
$|v_\theta|<\va$ throughout the flow, so there is no fast MHD
critical point.  The Alfv\'en critical point, however, does impose 
constraints on the possible values of the parameters $j$ and $e$.

At the \alf\ point where $\ma=1$, we use the streamline-invariant equations
(\ref{STREAMINV}) to evaluate $\omega$, $b$, $\ell$, and $\epsilon$ 
in terms of $\phi(\theta_\rma)\equiv \phi_\rma$.  Then, from 
equation (\ref{UPHIEQ}), a smooth solution requires $j\ell-\omega|_\rma=0$ 
so that
$\phi_\rma=j^{-(3-q)/4}$ must obtain.  Using 
the definition $\ams=m n b^2$, the reduced density at the \alf\ point
must be given by $n_\rma= j^{q/2}/m$.
The Bernoulli equation at the \alf\ point becomes
\begin{equation}
{d(\ams)\over d\theta}\bigg|_\rma =
{4\omega\phi'\over(3-q)\phi{\cal D}^{1/2}}\Bigg|_\rma,
\end{equation}
where $\cal D$ is defined in equation (\ref{DDEF}).
Finally, the Bernoulli and GS equations at the \alf\ point can be combined to 
yield the requirement that $\phi'_\rma$ must be a solution of 
\begin{equation}\label{GSALF}
1=
\left.{
(4\omega)^2\left[\omega^2 - 2e\epsilon + 
2\left({\ro\over\rl}\right)^3\sin\theta
-{h\phi\over(n b)^2}\left(\phi + {2\phi'\sin\theta\cos\theta\over 3-q}  \right)
\right]^2\over
{\cal D} 
\left\{
q\left[\omega^2 - 2e\epsilon + 2\left({\ro\over\rl}\right)^3\sin\theta\right]
+ 3\omega^2 - 2e\epsilon\right\}^2
}\right|_\rma.
\end{equation}

Thus, the values at the \alf\ point of $\phi$, $\omega$, $b$, $\ell$, 
$\epsilon$ and $n$
are determined from the initial conditions at the pole, while the required 
value of $\phi'_\rma$ is
given implicitly through equation (\ref{GSALF}) in terms of these known
parameters and the unknown position of the \alf\ point $\theta_\rma$. 
For arbitrary values of the input parameters $e$ and $j$, however, 
$\phi'$ at the point $\theta_\rma$ where $\phi=\phi_\rma$ will not
satisfy equation (\ref{GSALF}).   Thus, one of the input parameters 
becomes an eigenvalue which is determined by the condition that equation
(\ref{GSALF}) is indeed satisfied, so that the solution is regular at the
\alf\ point.  We evaluate the eigenvalue using the shooting method.
Specifically, we choose values of $q$ and $\ro/\rl$ 
(i.e. $e=(3/2)(\ro/\rl)^2$), and guess
a value of $\ra/\rl$ (i.e. $j=(\ra/\rl)^2$).  We then numerically
integrate from the pole up to the \alf\ point, defined by the criterion
$\phi=\phi_\rma$.  We use the error in equation (\ref{GSALF}) near the
\alf\ point to  
correct the guess for $\ra/\rl$, and repeat until a satisfactorily 
converged solution is obtained for the super-Alfv\'enic region 
(typical error is $\sim 10^{-6}-10^{-4}$ at the \alf\ point).
We can then step over the \alf\ point, and continue the integration
on the sub-Alfv\'enic side.

\subsection{Equatorial boundary conditions}\label{EQAAPP}

For integrations which start at the equator ($\theta=\pi/2$),
divergence of $n(\theta)$ (since $\vp\rightarrow 0$ in the cold-flow
limit) means we must analytically expand the relevant equations in
$\delta=\pi/2-\theta$.  For a  given value of $\ro/\rl$ (i.e. $e$), we
have $\phi(\pi/2)\equiv \phi_e=(\ro/\rl)^{-{3-q\over2}}$ from equation
(\ref{FLUX}), and from equation (\ref{STREAMINV}) we obtain the
equatorial values of the other streamline invariants.  Given an arbitrary value
for $\phi'(\pi/2)\equiv \phi'_e$, the lowest-order expansion of the
Bernoulli equation (\ref{SSBERN}) yields the approximation for the reduced
density function
\begin{eqnarray}
n(\delta)&=&(m b_e^2\delta)^{-1}
\left\{
h m^2 \left({\ro\over\rl}\right)^{-6}
\left[1 + \left({2\over 3-q} \right)^2\left({\phi'_e \over\phi_e}\right)^2
\right]\right.
+ \left.\left[\left({\ra\over\ro}\right)^2-1  \right]^2
\right\}^{1/2} 
\times\nonumber \\
& & \left[-1 +3\left({2\over 3-q} \right)^2
\left({\phi'_e \over\phi_e}\right)^2
\right]^{-1/2},
\end{eqnarray}
where here and in the following the ``$e$'' subscript denotes evaluation at
$\theta=\pi/2$.  From this equation, it is clear that solutions require
$(3-q)/(2\sqrt{3}) < |\phi'_e|/\phi_e$;  this condition is equivalent to the
well-known
requirement that streamlines leave the disk at an angle $>30^\circ$ with 
respect 
to the vertical (BP).

Next, given some choice for $\phi'_e$, we can expand the 
Grad-Shafranov equation (\ref{SSGSEQ}) to lowest order in $\delta$.  The 
result is linear equation which yields $\phi''_e$ in terms of $q$, $e$, $j$, 
and $\phi'_e$.  
For fixed $q$, we can identify a pair ($j$, $e$) 
which satisfies the requirements for a smooth transition at
the \alf\ point when approached from the pole, as described in \S\ref{ALFAPP}.
We can then iteratively solve for 
the value of $\phi'_e$ which allows an equatorially-initiated 
sub-Alfv\'enic solution to match
smoothly  onto the pole-initiated super-Alfv\'enic 
solution at the \alf\ point.  We 
generally proceed by choosing a value for $\ro/\rl$ and then searching 
for the unique values of $\ra/\rl$ and $\phi'_e$ which yield a smooth match at
the \alf\ point.  When sufficiently accurate values are determined, 
the pole-initiated and equator-initiated solutions can generally cross the
\alf\ point and overlap the solution on the far side, with very small
errors.

\clearpage


\clearpage

\end{document}